\documentclass[prb, superscriptaddress, notitlepage, twocolumn]{revtex4-1}
\usepackage{amsmath,amsthm,amsfonts,amssymb}
\usepackage{eucal}
\usepackage{tensor}
\usepackage{graphicx}
\usepackage{tabu}
\usepackage{indentfirst}
\usepackage{bbm}
\usepackage{grffile}
\usepackage[all,knot]{xy}
\usepackage{chngcntr}
\usepackage{floatrow}
\usepackage[caption=false]{subfig}
\usepackage[colorlinks, citecolor=blue, linkcolor=blue]{hyperref}
\usepackage[usenames, dvipsnames]{color}
\usepackage{enumitem,kantlipsum}
\usepackage{amsbsy}
\usepackage{multirow}
\usepackage{mathrsfs}
\xyoption{arc}

\newcommand{\A}{\mathcal{A}}
\newcommand{\M}{\mathcal{M}}

\newcommand{\D}{\mathcal{D}}

\newcommand{\mS}{\mathcal{S}}

\newcommand{\mfv}{\mathfrak{v}}
\newcommand{\mfp}{\mathfrak{p}}

\newcommand{\bta}{\text{Aut}^{\text{br}}_\otimes}
\newcommand{\brpic}{\text{BrPic}}
\newcommand{\pic}{\text{Pic}}
\newcommand{\Rep}{{\rm Rep}}

\DeclareMathOperator{\FPdim}{FPdim}

 \DeclareMathOperator{\Dim}{Dim} 
\DeclareMathOperator{\End}{End}  
\DeclareMathOperator{\Hom}{Hom} 
 
 \DeclareMathOperator{\Id}{Id}

\DeclareMathOperator{\Fun}{Fun}
\DeclareMathOperator{\Obj}{Obj}

\newcommand{\one}{\mathbf{1}}
\newcommand{\C}{\mathbb C}
\newcommand{\mC}{\mathcal{C}}
\newcommand{\mE}{\mathcal{E}}

\newcommand{\mZ}{\mathcal{Z}}
\newcommand{\mQ}{\mathcal{Q}}

\newcommand{\mfL}{\mathfrak{L}}
\newcommand{\mfC}{\mathfrak{C}}
\newcommand{\mfB}{\mathfrak{B}}
\newcommand{\mfr}{\mathfrak{r}}
\newcommand{\mfh}{\mathfrak{h}}
\newcommand{\mfD}{\mathfrak{D}}
\newcommand{\Z}{\mathbb Z}

\newcommand{\B}{\mathcal{B}}

\newcommand{\comments}[1]{}

\newcommand{\ket}[1]{|#1\rangle}

\renewcommand{\one}{\mathbf{1}}

\renewcommand{\D}{\mathcal{D}}
\renewcommand{\Z}{\mathbb{Z}}

\newcommand{\overbar}[1]{\mkern 2.3mu\overline{\mkern-2.3mu#1\mkern-2.3mu}\mkern 2.3mu}

\newtheorem{theorem}{Theorem}[section]

\newtheorem{corollary}[theorem]{Corollary}

\newtheorem{conj}[theorem]{Conjecture}

\theoremstyle{definition}

\newtheorem{remark}[theorem]{Remark}

\newtheorem{definition}[theorem]{Definition}
\newtheorem{defthm}[theorem]{Definition/Theorem}

\begin{document}

\title{On Defects Between Gapped Boundaries in Two-Dimensional\\Topological Phases of Matter}
\author{Iris Cong}
\affiliation{Department of Computer Science, University of California, Los Angeles, CA 90095, U.S.A.} 
\affiliation{Microsoft Station Q, University of California, Santa Barbara, CA 93106-6105 U.S.A.}
\author{Meng Cheng}
\affiliation{Microsoft Station Q, University of California, Santa Barbara, CA 93106-6105 U.S.A.}
\affiliation{Department of Physics, Yale University, New Haven, CT 06520-8120, U.S.A.} 
\author{Zhenghan Wang}
\affiliation{Microsoft Station Q, University of California, Santa Barbara, CA 93106-6105 U.S.A.}
\affiliation{Department of Mathematics, University of California, Santa Barbara, CA 93106-6105 U.S.A.}

\begin{abstract}
Defects between gapped boundaries provide a possible physical realization of projective non-abelian braid statistics. A notable example is the projective Majorana/parafermion braid statistics of boundary defects in fractional quantum Hall/topological insulator and superconductor heterostructures. In this paper, we develop general theories to analyze the topological properties and projective braiding of boundary defects of topological phases of matter in two spatial dimensions. We present commuting Hamiltonians to realize defects between gapped boundaries in any $(2+1)D$ untwisted Dijkgraaf-Witten theory, and use these to describe their topological properties such as their quantum dimension. By modeling the algebraic structure of boundary defects through multi-fusion categories, we establish a bulk-edge correspondence between certain boundary defects and symmetry defects in the bulk. Even though it is not clear how to physically braid the defects, this correspondence elucidates the projective braid statistics for many classes of boundary defects, both amongst themselves and with bulk anyons. Specifically, three such classes of importance to condensed matter physics/topological quantum computation are studied in detail: (1) A boundary defect version of Majorana and parafermion zero modes, (2) a similar version of genons in bilayer theories, and (3) boundary defects in $\mfD(S_3)$.
\end{abstract}

\maketitle
{\hypersetup{linkcolor=black}
\tableofcontents}


\section{Introduction}
\label{sec:intro}

\subsection{Motivations}
\label{sec:motivations}

The Ising anyon $\sigma$ with quantum dimension $\sqrt{2}$ is arguably the most famous non-abelian object.  Incarnations of the Ising anyon include the Majorana zero mode at the ends of Majorana nanowires, defects between the rough and smooth gapped boundaries of the toric code, and the symmetry defect associated with the $\mathbb{Z}_2$ electric-magnetic symmetry of the toric code (see Ref. \onlinecite{DFN} and references therein).  The Majorana zero mode of nanowires can be understood either as a defect of fermions associated to the fermion parity or as defects between different gapped boundary types (a.k.a. {\it boundary defects}) of topological insulators \cite{FuKa}.  This circle of phenomena around the Majorana zero mode has been generalized to the parafermion zero mode~\cite{clarke2013, cheng2012,lindner2012}.  In this paper, we discover that the correspondence between some boundary defects and $\mathbb{Z}_2$ symmetry defects is a general phenomenon.  This bulk-boundary correspondence explains the projective non-abelian statistics of parafermion zero modes, and allows us to define projective statistics of boundary defects between certain gapped boundaries.  We also generalize the correspondence of genons and boundary defects in abelian theories to general theories including non-abelian ones.

Boundary defects such as parafermion zero modes and genons are pursued as a mechanism to generate non-abelian objects from abelian materials, whose existence is more certain.  Following Refs. \onlinecite{Cong16a,Cong16b}, we provide local commuting Hamiltonians for realizing defects between gapped boundaries for the untwisted Dijkgraaf-Witten theories.  We obtain a group-theoretical description of boundary defects and derive simple formulas for their quantum dimensions.  

The physical realizations of parafermion zero modes in Refs. \onlinecite{clarke2013, cheng2012,lindner2012} are experimentally accessible.  Direct analogues of those parafermion zero modes are realized by our models using defects between different gapped boundaries of the $\mathbb{Z}_p$ toric code. Different gapped boundaries of the $\mathbb{Z}_p$ toric code are abstractions of the superconducting segment and magnetically insulating segment of the boundary of a fractional topological insulator disk, where the parafermion zero modes emerge at the interfaces of different segments. Projective braid statistics of these parafermions zero modes are computed in Refs. \onlinecite{clarke2013, cheng2012,lindner2012} and implemented by a measurement only protocol.  The boundary defects in the toric code ($p=2$) is part of the surface code theory, and building a quantum computer based on surface codes is experimentally pursued \cite{Fowler12, BK+14}.

\subsection{Previous Works}
\label{sec:previous-works}

Boundary defects in abelian topological phases have received much attention because their physical realization provides a new way to \lq\lq engineer" non-abelian objects in an abelian parent state, and by now, a relatively complete physical understanding has been achieved. The simplest non-abelian boundary defect, emerging between gapped boundaries of the $\mathbb{Z}_2$ toric code, was first considered in Ref. \onlinecite{Bravyi98} in the context of quantum codes. More recently, similar non-abelian defects have been proposed to be realized in fractional quantum Hall systems (e.g. FQH-superconductor heterostructures~\cite{clarke2013, cheng2012,lindner2012}, ``genons'' in bilayer FQH systems~\cite{Barkeshli11, Bark13a}). A general discussion of such boundary defects, based on the classification of gapped boundaries in abelian phases~\cite{Levin13,LWW,WW,Hung15a,Hung15b}, can be found in Refs. \onlinecite{Bark13b, Bark13c, Kapustin14}. 

The Hamiltonian for boundary defects in the Dijkgraaf-Witten theory based on a finite group $G$ is a generalization of the gapped boundary Hamiltonians presented in Refs. \cite{Kitaev97,Bombin08,Beigi11}. In particular, we generalize the Hamiltonians of Ref. \cite{Bombin08} to gapped boundaries given by non-normal subgroups of $G$, and we generalize Ref. \cite{Beigi11} by considering gapped boundaries of arbitrary shape on the lattice.

The topological degeneracy associated with multiple boundary defects can also be exploited for topological quantum information processing, for which it is crucial to be able to implement protected gate operations on the degeneracy subspace.  Although we do not know how to  physically ``braid'' the defects, the idea of measurement-only braiding can be adopted to generate analogues of braiding transformations for boundary defects~\cite{clarke2013, lindner2012, Bonderson13}.

The higher category theory necessary for modeling boundary defects can be found in Ref. \onlinecite{Etingof15}, and modeling of boundary defects using functors between module categories is proposed in Ref. \onlinecite{KitaevKong}.

\subsection{Main Results}
\label{sec:main-results}

Our main contributions include a general understanding of how the ground state  degeneracy of many boundary defects can support non-abelian braid statistics projectively, and commuting local projector Hamiltonians to realize boundary defects between gapped boundaries in any untwisted Dijkgraaf-Witten theory.  In particular, this explains the projective braid statistics of parafermion zero modes realized by FQH-superconductor heterostructures, which are computed in Refs. \onlinecite{clarke2013, cheng2012,lindner2012}.

The detailed contents of the paper are as follows. As a motivational example, in Section \ref{sec:majorana-parafermion-hamiltonian}, we present commuting projector Hamiltonians that generate boundary defects in Kitaev's $\Z_p$ toric code \cite{Kitaev97} ($p \geq 2$ is an arbitrary prime).  Such boundary defects realize the same projective braid statistics of the parafermion zero modes ($p>2$) and Majorana zero modes ($p=2$).  In Section \ref{sec:hamiltonian-realization}, we provide the generalization of these commuting projector Hamiltonians to generate the defect between any two gapped boundaries given by subgroups $K_1, K_2 \subseteq G$, in any untwisted Dijkgraaf-Witten theory.  We find that the simple defect types are given by the set

\begin{equation}
\{
(T,R): T \in K_1 \backslash G / K_2, \text{ } R \in (K_1, K_2)^{r_T}_{\text{ir}} 
\}.
\end{equation}

\noindent
where $(K_1, K_2)^{r_T} = K_1 \cap r_T K_2 r_T^{-1}$ for some representative $r_T \in T$. Furthermore, the quantum dimension of the simple defect is given by:\footnote{We use the notation $\text{FPdim}$ to denote the quantum dimension of the anyon corresponding to the label $(T,R)$, to contrast with the notation $\Dim$ for the usual dimension of a representation.}

\begin{equation}
\text{FPdim}(T,R) = \frac{\sqrt{|K_1| |K_2|}}{|(K_1, K_2)^{r_T}|} \cdot \Dim(R).
\end{equation}

In Section \ref{sec:algebraic-theory}, we model boundary defects using category theory. Because the category of defects between two specific gapped boundary types does not have the proper tensor product (fusion) structure, we instead consider all possible boundary excitations and boundary defects of the topological phase together in the multi-fusion category $\mfC$. In this category, we can compute the topological properties of boundary excitations or defects, such as quantum dimensions and fusion rules. We then use this model in Section \ref{sec:bulk-edge-correspondence} to extend the bulk-to-boundary condensation functor \cite{Kong15,Cong16a,Cong16b} to a correspondence between bulk symmetry defects and boundary defects. We find that bulk symmetry defects, when dragged to a gapped boundary, will condense to boundary defects, just as bulk anyons condense to boundary excitations. In particular, when a symmetry defect carrying flux $g \in G$ is brought to a gapped boundary line, it causes one side of the boundary line to undergo a $g$ action (Figs. \ref{fig:crossed-condensation-a}, \ref{fig:crossed-condensation-b}); this turns the symmetry defect into a defect between two gapped boundary types. 

This far-reaching and general understanding underpins the projective braid statistics supported by the topological degeneracy of boundary defects, and provides the theoretical foundation for topological quantum computation using boundary defects. Specifically, we find in Section \ref{sec:braiding} that the projective braiding of boundary defects between two gapped boundary types is be determined by the $G$-crossed braiding of their bulk symmetry defect counterparts, if the given boundary types are related by the bulk $G$ symmetry action. This explains the braiding of boundary defects among themselves, with boundary excitations, and with bulk anyons.

In Section \ref{sec:examples}, three important families of examples in condensed matter physics and topological quantum computation are studied in detail: (1) A boundary defect version of Majorana and parafermion zero modes, (2) a similar version of genons in bilayer theories, and (3) boundary defects in $\mfD(S_3)$.

\subsection{Notations}


Throughout the paper, all algebras and tensor categories are over the complex numbers $\C$.  All fusion and modular tensor categories are unitary.  Unitary fusion categories are spherical. The Drinfeld center of a unitary fusion category $\mC$ is denoted $\mZ(\mC)$. In the group-theoretical case where $\mC = \text{Vec}_G$ for some finite group $G$, we write $\mfD(G) = \mZ(\text{Vec}_G) = \mZ(\text{Rep}(G)) = \Rep(D(G))$, where $D(G)$ is the quantum double of $G$, $\text{Vec}_G$ is the category of $G$-graded vector spaces, and $\Rep(G)$ is the representation category of $G$. In general, if $G$ is any finite group, we denote the set of irreducible representations of $G$ by $(G)_{\text{ir}}$.

In this paper, gapped boundaries will always be oriented so that the bulk of the topological phase is on the left hand side when traversing the boundary; this allows us to consider gapped boundaries as indecomposable left module categories. In general, the defects between gapped boundaries will be marked with an $X$ in figures, and bulk symmetry defects will be marked with a $Y$ in figures.

\subsection{Acknowledgment}
The authors thank Maissam Barkeshli, Shawn Cui, and Cesar Galindo for answering many questions. I.C. would like to thank Michael Freedman and Microsoft Station Q for hospitality in hosting the summer internship and visits during which this work was done. M.C. thanks Chao-Ming Jian for collaborations on related topics. Z.W. is partially supported by NSF grant DMS-1411212.

\vspace{2mm}

\section{Majorana and Parafermion Zero Modes as Boundary Defects}
\label{sec:majorana-parafermion-hamiltonian}

In this section, we present a Hamiltonian which generates boundary defects in Kitaev's $\Z_p$ toric code \cite{Kitaev97}, where $p \geq 2$ is an arbitrary prime. Such boundary defects are one incarnation of the parafermion zero modes, and in the special case where $p = 2$, they become Majorana zero modes.

\begin{figure}
\centering
\includegraphics[width = 0.8\textwidth]{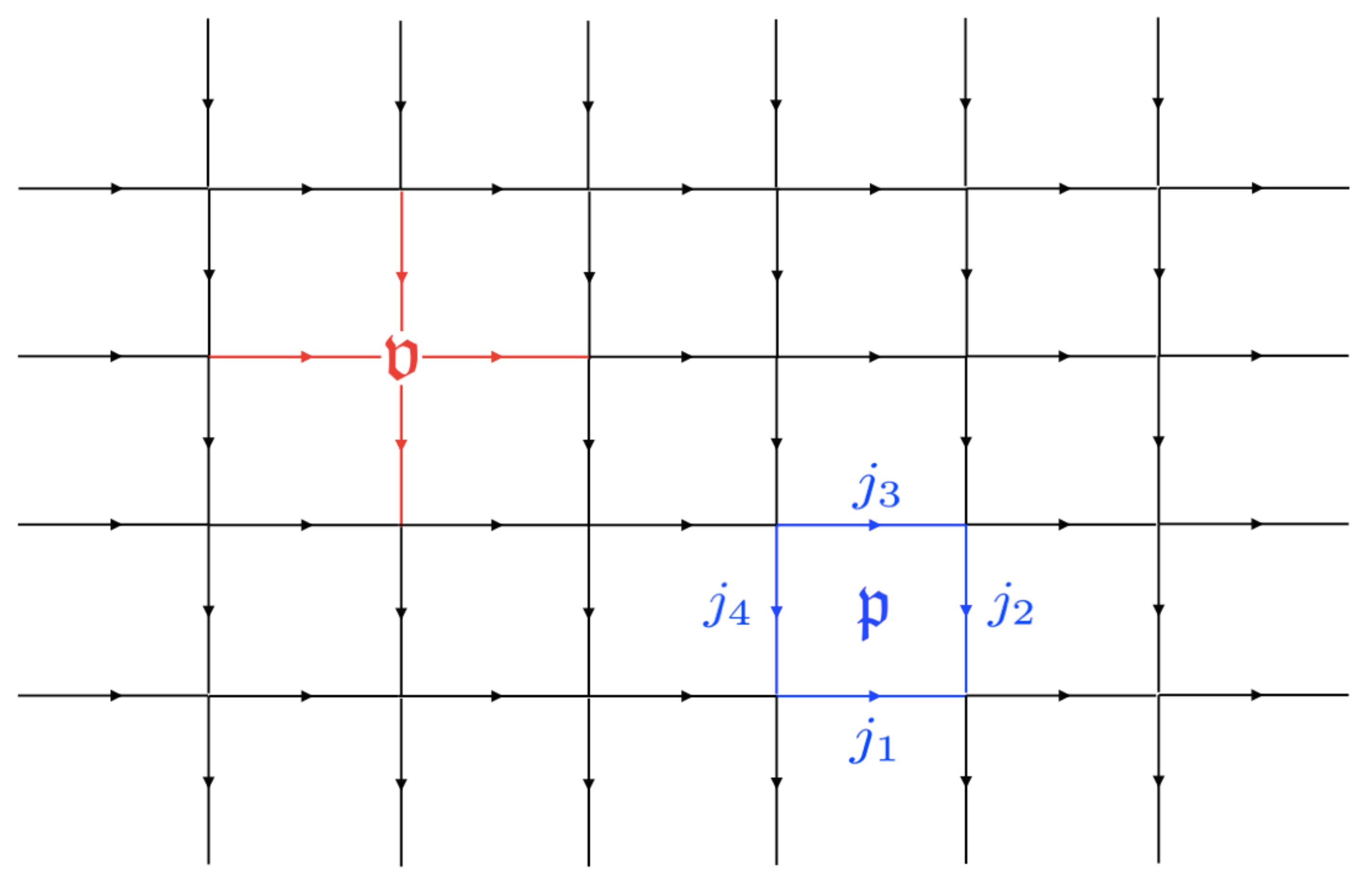}
\caption{Square lattice for the toric code Hamiltonian. For the example vertex $\mfv$ (plaquette $\mfp$), the edges in $\text{star}(\mfv)$ ($\text{boundary}(\mfp)$) are highlighted in red (blue).}
\label{fig:square-lattice}
\end{figure}

Let $\mathcal{H}$ be a $p$-dimensional Hilbert space with orthonormal basis $\ket{0}$, $\ket{1}$, ... $\ket{p-1}$. Given a square lattice as in Fig. \ref{fig:square-lattice}, on each edge of the lattice, we place a qudit taking values in $\mathcal{H}$. Define the generalized Pauli matrices $X_p$ and $Z_p$ such that

\begin{equation}
X_p \ket{i} = \ket{i+1} \qquad
Z_p \ket{i} = \omega^{p} \ket{i}
\end{equation}

\noindent
where addition is done modulo $p$ and $\omega = e^{2\pi i/p}$ is the $p^{\text{th}}$ root of unity. Because $G = \Z_p$ is cyclic, Kitaev's Hamiltonian \cite{Kitaev97} for the bulk of the $\Z_p$ toric code is (up to a constant):

\begin{equation}
\label{eq:toric-code-hamiltonian-bulk}
H_{\text{bulk}} = - \sum_\mfv \sum_{i=0}^{p-1} \prod_{j \in \text{star}(\mathfrak{v})} X_p^i(j) - \sum_\mfp \sum_{i=0}^{p-1} \prod_{j \in \text{bd}(\mathfrak{p})} Z_p^i(j).
\end{equation}

\noindent
In the above equation, we sum over all vertices $\mfv$ and all plaquettes $\mfp$ in the lattice. Fig. \ref{fig:square-lattice} shows one example of $\text{star}(\mfv)$ and $\text{bd}(\mfp)$.   The operators $X_p(j)$ and $Z_p(j)$ are extensions of $X_p$ and $Z_p$, which act on the qudits at the edge $j$, by tensoring the identities on the other qudits.

The hole Hamiltonian defined in Refs. \onlinecite{Cong16a,Cong16b} will allow us to generate gapped boundaries and boundary defects. In general, a gapped boundary type in a quantum double model $\mfD(G)$ is determined by a subgroup $K \subseteq G$ up to conjugation and a 2-cocycle of $K$. Since $p$ is a prime, the group $\Z_p$ has exactly two subgroups, namely the trivial subgroup and $\Z_p$ itself, and the cohomology class is trivial in both cases. In this section, we consider $K = \Z_p$; the other case is almost identical.

With this setup, the Hamiltonian for the hole as discussed in Refs. \onlinecite{Cong16a,Cong16b} is:

\begin{equation}
\label{eq:toric-code-hamiltonian-hole}
H_{\text{hole}} = - \sum_\mfv \sum_{i=0}^{p-1} \prod_{j \in \text{star}(\mathfrak{v})} X_p^i(j) - \sum_{\text{edges } e} \sum_{i=0}^{p-1} X_p^i(e).
\end{equation}

\begin{figure}
\centering
\includegraphics[width = \textwidth]{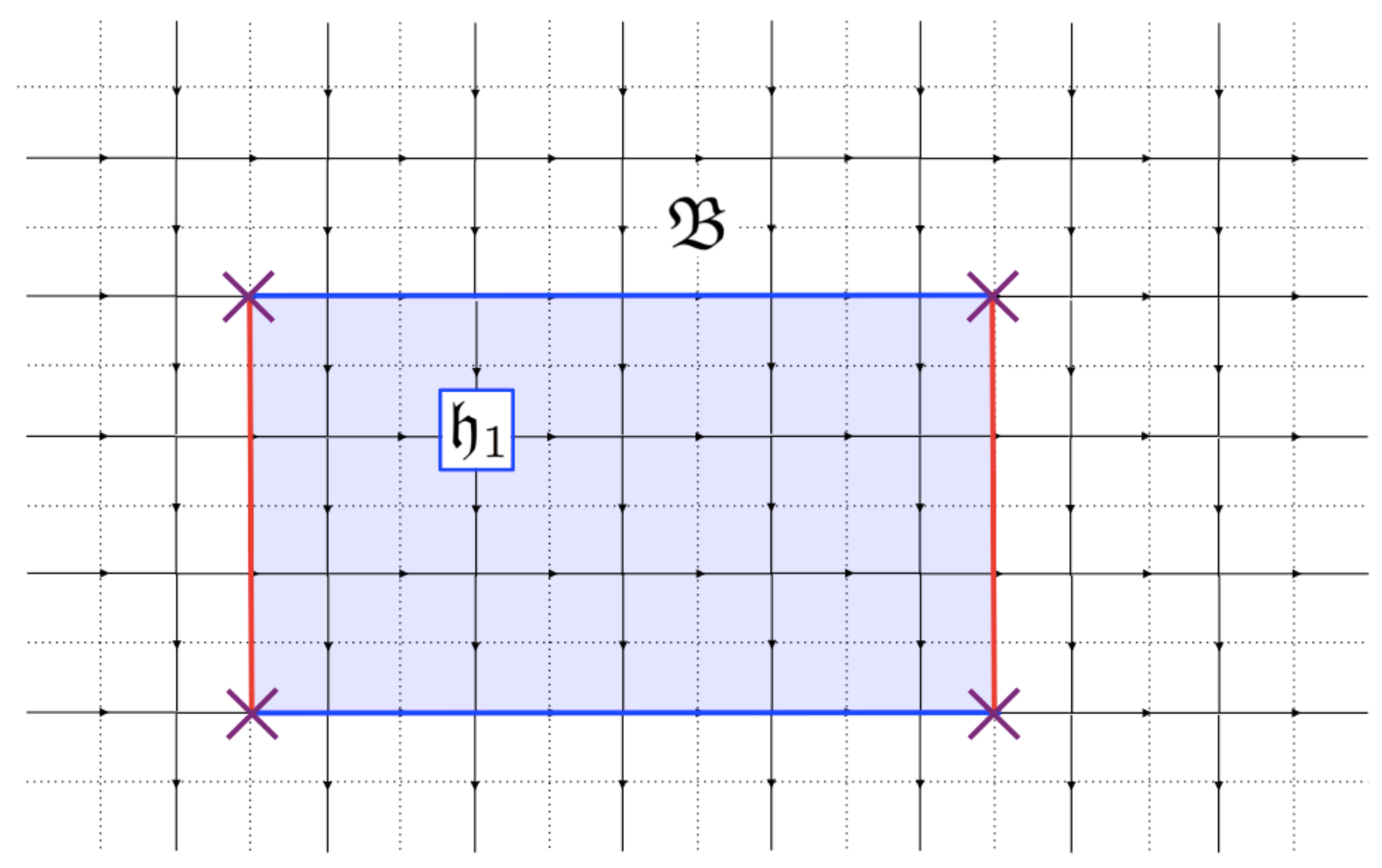}
\caption{Definition of the Hamiltonian for parafermion zero modes. Note the hole $\mfh_1$ includes all direct and dual vertices on its border (the red/blue boldfaced lines on the lattice) but NOT the edges on the border. The red lines are rough boundaries, and the blue lines are smooth boundaries. Parafermion zero modes are created at the four corners of $\mfh_1$ (purple crosses).}
\label{fig:parafermion-hamiltonian-def}
\end{figure}

Finally, boundary defects are created through the following commuting Hamiltonian:

\begin{equation}
H_{\text{parafermion}} = H_{\text{bulk}} (\mfB) + H_{\text{hole}} (\mfh_1)
\end{equation}

\begin{figure}
\centering
\includegraphics[width = \textwidth]{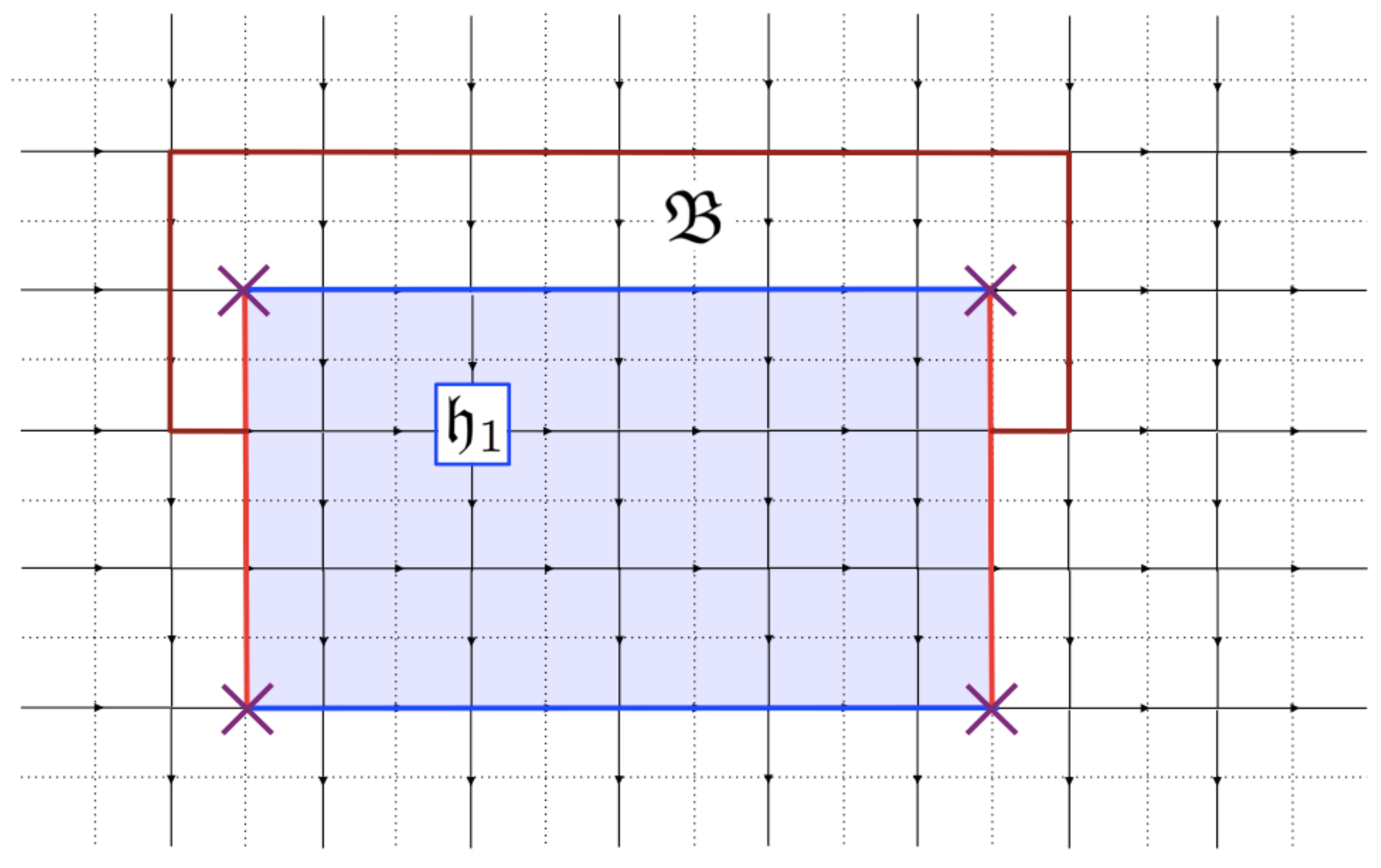}
\caption{Ground state degeneracy of the parafermion Hamiltonian. The ground state of $H_{\text{parafermion}}$ is $p$-fold degenerate, with a basis given by $\{ \ket{0}, \ket{1}, ... \ket{p-1} \}$, where $\ket{0}$ is the ground state of the bulk Hamiltonian $H_{\text{bulk}}$, and $\ket{i}$ is obtained from $\ket{0}$ by applying the operator $X_p^i$ to all qudits along the dark red line.}
\label{fig:parafermion-hamiltonian-degeneracy}
\end{figure}

Here, the regions $\mfB$ and $\mfh_1$ are as defined in Fig. \ref{fig:parafermion-hamiltonian-def}. The hole $\mfh_1$ includes all vertices and plaquettes on its border (the red/blue boldfaced lines on the lattice) but NOT the edges on the border. The red lines are rough boundaries, and the blue lines are smooth boundaries. The Hamiltonian creates boundary defects (the parafermion zero modes) at the four corners of $\mfh_1$ indicated by purple crosses in the figure. The fact that these defects are parafermion zero modes is verified by the ground state degeneracy: the ground state with four defects is degenerate with a basis $\{ \ket{0}, \ket{1}, ... \ket{p-1} \}$, where $\ket{0}$ is the ground state of the bulk Hamiltonian $H_{\text{bulk}}$, and $\ket{i}$ is the state $\ket{0}$ with the operator $X_p^i$ applied to all qudits along the dark red line in Fig. \ref{fig:parafermion-hamiltonian-degeneracy}. This $p$-fold degeneracy is consistent with the fusion rules of the parafermion zero modes. In Section \ref{sec:majorana-parafermion}, we also compute the projective braiding statistics of these defects to further illustrate that they correspond to parafermion zero modes.

\begin{remark}
If we had used the special case $p=2$ throughout this section, the bulk theory would be Kitaev's $\Z_2$ toric code, and all the sums over $i$ are unnecessary in Eqs. (\ref{eq:toric-code-hamiltonian-bulk}) and (\ref{eq:toric-code-hamiltonian-hole}) and can be removed.
\end{remark}

\vspace{2mm}
\section{Hamiltonian Realization of Boundary Defects}
\label{sec:hamiltonian-realization}

In the previous section, we presented a Hamiltonian to realize boundary defects in the $\Z_p$ toric code, which corresponded to Majorana/parafermion zero modes.
We now consider the more general case: starting from any untwisted Dijkgraaf-Witten theory based on a finite group $G$, we write down a commuting projector Hamiltonian for boundary defects. In particular, our Hamiltonian can realize the defect between any two gapped boundaries given by subgroups $K_1, K_2 \subseteq G$.

\subsection{Hamiltonians for the bulk and gapped boundaries}
\label{sec:hamiltonian-background}

In writing down our commuting projector Hamiltonian for defects, we will combine various projector terms from Kitaev's original quantum double models \cite{Kitaev97} and some extensions of these as developed in Ref. \onlinecite{Bombin08} to enhance the gapped boundary Hamiltonian of Refs. \onlinecite{Cong16a,Cong16b}. Let us briefly review these projectors and Hamiltonians that will be essential for the rest of the section (for more details, see Sections 2.1-2.3 in Refs. \onlinecite{Cong16a,Cong16b})

As in the previous section, let us start with a square lattice as shown in Fig. \ref{fig:square-lattice}\cite{Note1}. Since we now start with an arbitrary finite group $G$ which may be non-abelian, we note that the edges must be oriented; for convenience, we orient them all down and to the right.

As before, a qudit is placed on each edge of the lattice, whose Hilbert space is spanned by $\{\ket{g}: g \in G\}$, so the total Hilbert space for the quantum system is $\mathcal{L}=\otimes_{e}\mathbb{C}[G]$. Instead of the generalized Paulis $X_p,Z_p$, we must now use new projector terms which correspond to a multiplication/comultiplication in the group algebra $\mathbb{C}[G]$:

\begin{equation}
\label{eq:L}
L^{g_0}_+ \ket{g} = \ket{g_0g} \qquad
L^{g_0}_- \ket{g} = \ket{gg_0^{-1}}
\end{equation}
\begin{equation}
\label{eq:T}
T^{h_0}_+ \ket{h} = \delta_{h_0,h}\ket{h}
\qquad 
T^{h_0}_- \ket{h} = \delta_{h_0^{-1},h}\ket{h}
\end{equation}

\noindent
where $\delta_{i,j}$ is the Kronecker delta function. These projector terms are defined for all elements $g_0,h_0 \in G$. Using these projections, local gauge transformations and magnetic charge operators are defined as follows, on each vertex $\mfv$ and plaquette $\mfp$ \cite{Kitaev97}:

\begin{equation}
\label{eq:kitaev-vertex-g-term}
A^{g}(\mfv,\mfp) = A^{g}(\mfv) = \prod_{j \in \text{star}(\mfv)} L^g(j,\mfv)
\end{equation}

\begin{equation}
\label{eq:kitaev-plaquette-h-term}
B^h(\mfv,\mfp) = \sum_{h_1 \cdots h_k = h} \prod_{m=1}^k T^{h_m}(j_m, \mfp)
\end{equation}

Here, $j_1, ..., j_k$ are the boundary edges of the plaquette $\mfp$ in counterclockwise order, starting from the vertex $\mfv$ (see Fig. \ref{fig:square-lattice}), and $L^g$ and $T^h$ are defined as follows: if $\mfv$ is the origin of the directed edge $j$, $L^g(j,\mfv) = L^g_-(j)$, otherwise $L^g(j,\mfv) = L^g_+(j)$; if $\mfp$ is on the left (right) of the directed edge $j$, $T^h(j,\mfp) = T^h_-(j)$ ($T^h_+(j)$) \cite{Kitaev97}. Two linear combinations of $A^g$ and $B^h$ are used to define Kitaev's Hamiltonian for the bulk of the Dijkgraaf-Witten theory:

\begin{equation}
\label{eq:kitaev-vertex-term}
A(\mfv) = \frac{1}{|G|} \sum_{g \in G} A^g(\mfv,\mfp) \qquad
B(\mfp) = B^1(\mfv,\mfp).
\end{equation}

\noindent
The Hamiltonian for the Kitaev model is then defined as

\begin{equation}
\label{eq:kitaev-hamiltonian}
H_G = -\sum_\mfv A(\mfv) - \sum_\mfp B(\mfp).
\end{equation}

\noindent
It is well known that all terms in the Hamiltonian $H_G$ commute with each other, so that the spectrum of $H_G$ is always gapped, giving a ``topological'' encoding of information in the ground state manifold.

On a surface with trivial topology like the sphere or an infinite plane, the ground state of $H_G$ is non-degenerate. In Refs. \onlinecite{Cong16a,Cong16b}, to introduce degeneracy into the ground state, we extended this Hamiltonian to surfaces with boundary, while still only using local commuting projector terms. In general, such a {\it gapped boundary} in the Dijkgraaf-Witten theory based on $G$ is given by a subgroup $K \subseteq G$ up to conjugation and a 2-cocycle $\phi \in H^2(K,\C^\times)$. We considered the cases where $\phi = 1$, but a generalization is straightforward. In these cases, the Hamiltonian was defined with some new projector terms (first introduced in Ref. \onlinecite{Bombin08} for the special case of $K$ normal in $G$).

\begin{equation}
\label{eq:LK}
L^K(e) := \frac{1}{|K|} \sum_{k \in K} (L^k_+ (e) + L^k_- (e)),
\end{equation}

\begin{equation}
\label{eq:TK}
T^K(e) := \sum_{k \in K} T^k_+(e)
\end{equation}

\noindent
The operators $L^k_{+,-}$ and $T^k_+$ are defined in Eqs. (\ref{eq:L}) and (\ref{eq:T}).
The choice of using only $T_+$ in the definition of $T^K$ is arbitrary, as using only $T_-$ would yield the same operator.


A commuting Hamiltonian for the hole of a gapped boundary was then defined as:

\begin{equation}
\label{eq:bd-hamiltonian-K}
H^K_G =  
 - \sum_e (T^K(e)+L^K(e))
\end{equation}

\begin{figure}
\centering
\includegraphics[width = 0.8\textwidth]{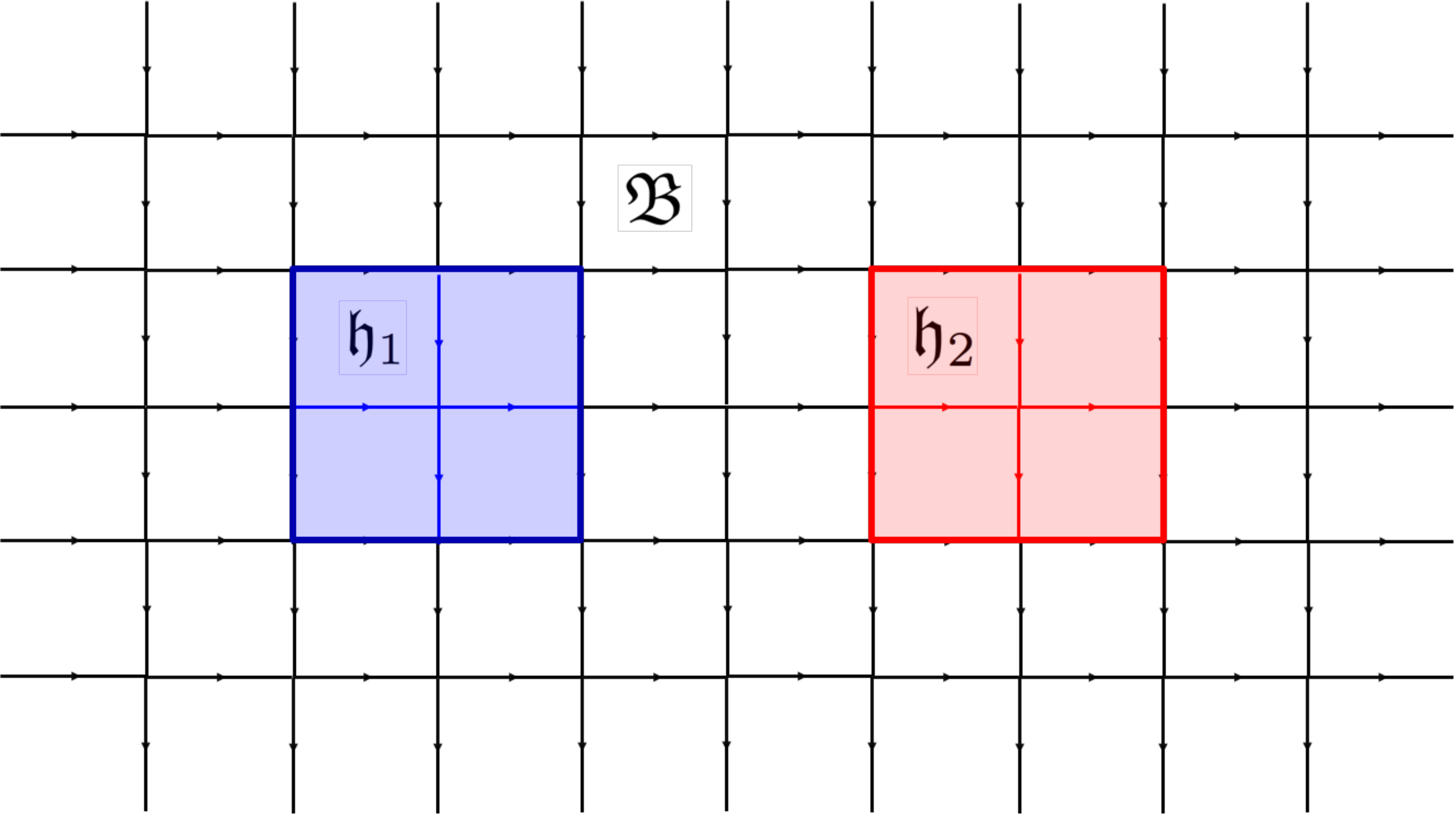}
\caption{Example: defining $H_{\text{G.B.}}$ in the case of two holes on an infinite lattice.}
\label{fig:boundary-hamiltonian}
\end{figure}

In the above Hamiltonian, 
the sum runs over all edges within and along the boundary. Using this, a commuting Hamiltonian was defined to realize $n$ gapped boundaries given by subgroups $K_1, ... K_n$ (see Fig. \ref{fig:boundary-hamiltonian} for example):

\begin{equation}
\label{eq:gapped-bds-hamiltonian}
H_{\text{G.B.}} = H_G(\mathfrak{B}) + \sum_{i=1}^{n} H^{K_i}_{G}(\mathfrak{h}_i).
\end{equation}

\noindent
Here, $H^{K_i}_{G}(\mathfrak{h}_i)$ indicates that the Hamiltonian $H^{K_i}_{G}$ is acting on all edges of the hole $\mathfrak{h}_i$, and similarly for $H_G(\mathfrak{B})$. 


\subsection{The commuting projector Hamiltonian for boundary defects}
\label{sec:defect-hamiltonian}

\begin{figure}
\centering
\includegraphics[width = \textwidth]{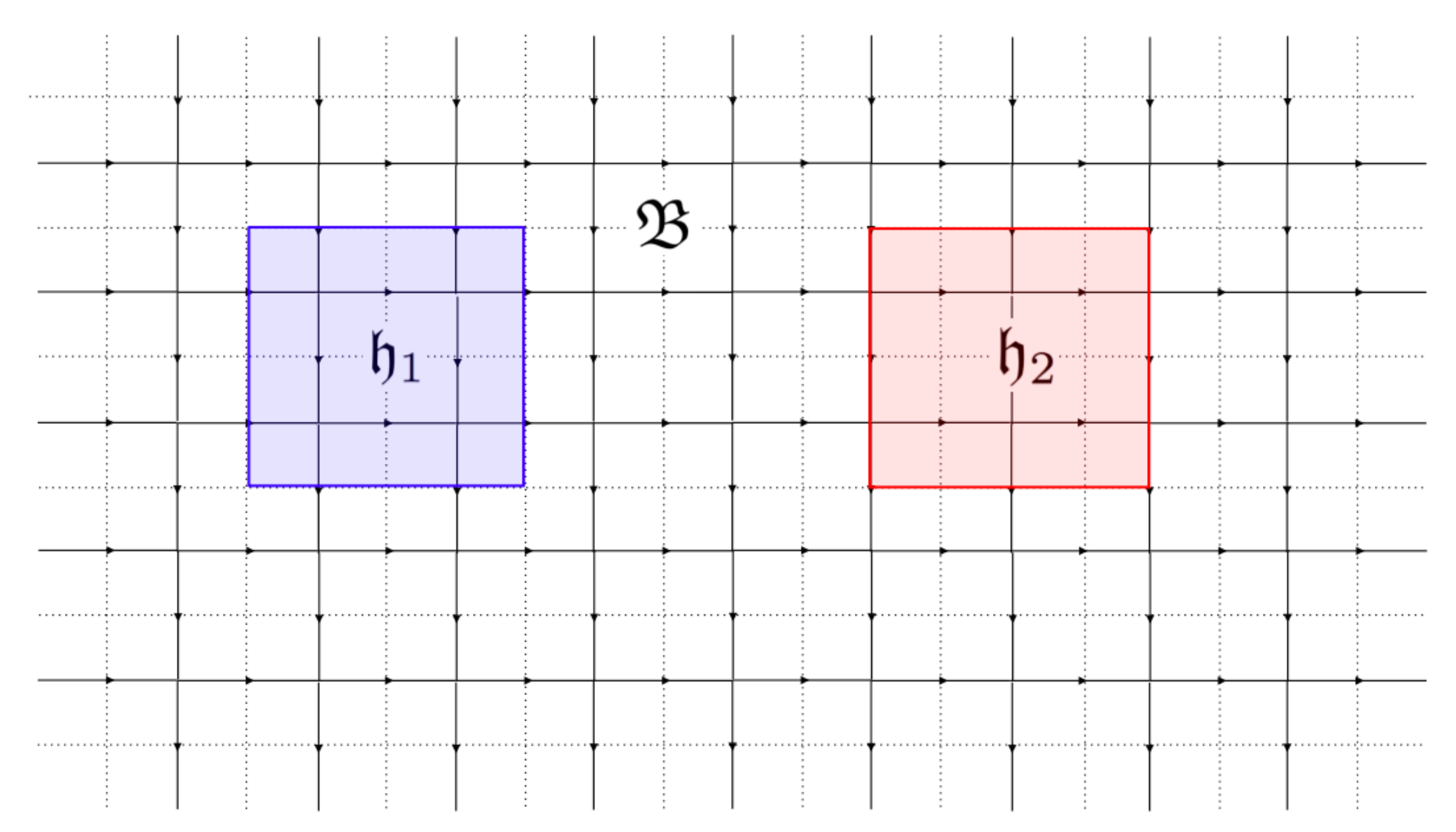}
\caption{Definition of the Hamiltonian $H_{\text{G.B.}}$, in cases where some (e.g. $\mathfrak{h}_2$) or all (e.g. $\mathfrak{h}_1$) of the hole's sides lie on the dual lattice. For $\mathfrak{h}_2$, four boundary defects are created, one at each corner.}
\label{fig:boundary-hamiltonian-3}
\end{figure}

In general, there are multiple ways to create defects between different boundary types. The first and simplest way is a direct generalization of the Majorana/parafermion zero mode Hamiltonian of Section \ref{sec:majorana-parafermion-hamiltonian}:

In the previous section, we showed how to create gapped boundaries, where the sides of the corresponding holes $\mfh_i$ lie on the lattice. More generally, any lattice in the plane has a {\it dual lattice} formed by switching vertices/plaquettes (e.g. dotted lines in Fig. \ref{fig:boundary-hamiltonian-3}). As discussed in Refs. \onlinecite{Cong16a,Cong16b}, we can create also holes where some or all of the sides lie on the dual lattice, such as $\mathfrak{h}_1$ and $\mathfrak{h}_2$, respectively, in Fig. \ref{fig:boundary-hamiltonian-3}. For both cases, the Hamiltonian $H^{K_i}_{G}$ acts on edges within and along the boldfaced boundary of the square. In general, the properties of holes completely on the dual lattice such as $\mathfrak{h}_1$ are almost the same as those of holes on the direct lattice; the only difference is up to an electric-magnetic symmetry in the model. However, holes such as $\mathfrak{h}_2$ that are partially on the dual lattice are of more interest. In this case, the hole is essentially associated with two different boundary types (e.g. two different subgroups $K_1, K_2 \subseteq G$), related to each other by this symmetry; the Hamiltonian $H_{\text{G.B.}}$ then creates defects between these two boundary types at each corner of the hole where a lattice boundary line meets a dual lattice boundary line. For the special case of $G = \Z_p$, the two subgroups were $K_1 = \{0\}$ and $K_2 = \Z_p$, and the corresponding defect was the parafermion zero mode.

However, the above procedure only allows us to create very special kinds of defects: the two boundary types involved must be the direct/dual lattice boundary types corresponding to a common subgroup $K \subseteq G$. More generally, we would like to create defects between arbitrary boundary types given by subgroups $K_1, K_2 \subseteq G$. This can be done by defining a new commuting Hamiltonian $H_{\text{dft}}$.

\begin{figure}
\centering
\includegraphics[width = \textwidth]{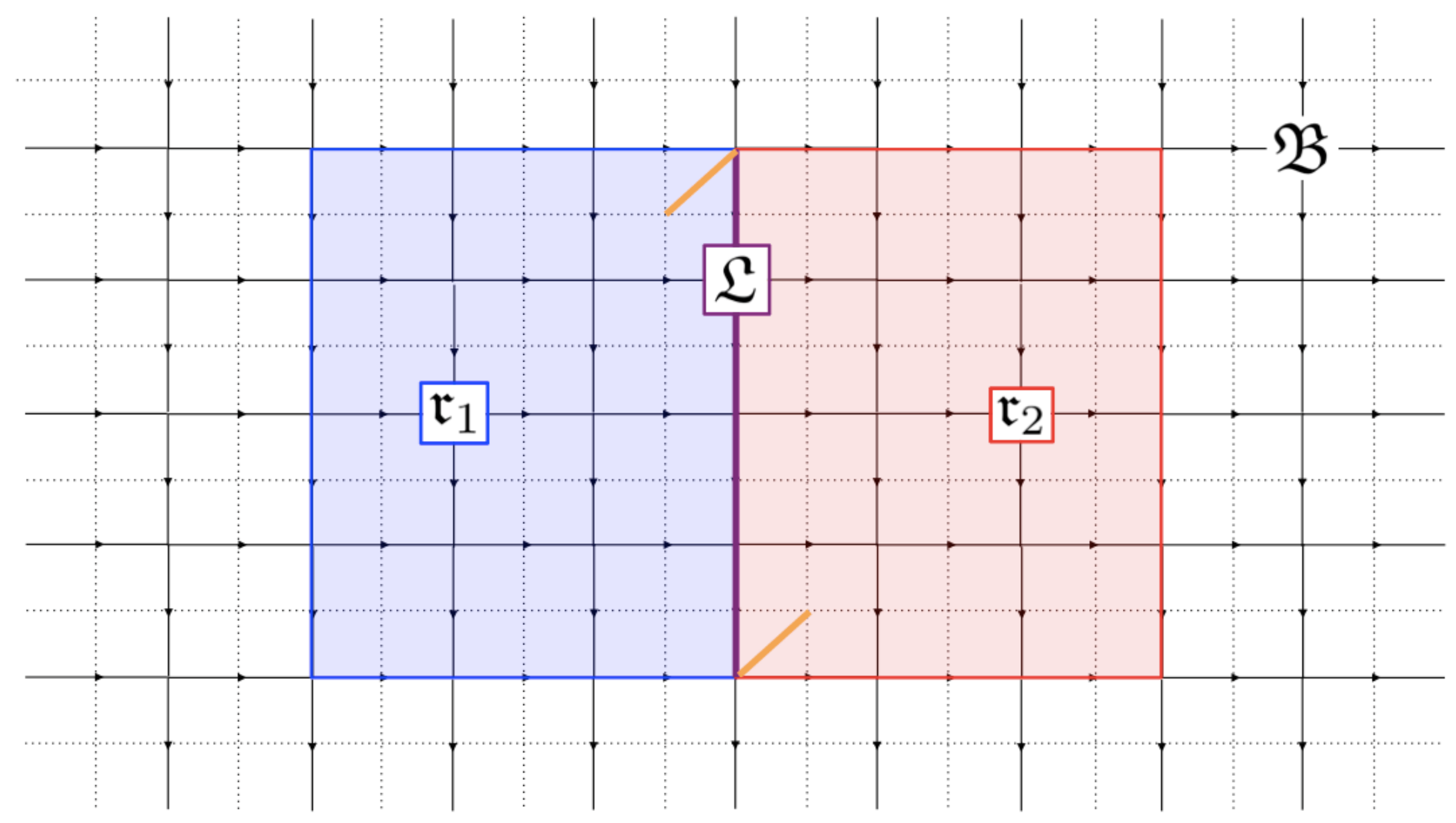}
\caption{Definition of the Hamiltonian $H_{\text{dft}}$. The line $\mathfrak{L}$ consists of all vertices and edges along the purple line which divides the blue and red regions. The region $\mathfrak{r}_1$ consists of all vertices, plaquettes, and edges within and along the blue shaded rectangle. $\mathfrak{r}_2$ is defined similarly. As before, the bulk $\mfB$ denotes all other vertices, edges, and plaquettes (i.e. the black and white region). The resulting defects live on the orange cilia.} 
\label{fig:defect-hamiltonian}
\end{figure}

To define this new Hamiltonian, let us first consider a picture such as Fig. \ref{fig:defect-hamiltonian}. Suppose we would like to create two defects, one at each endpoint of the line $\mfL$. Specifically, we would like the boundary type to the left of $\mfL$ (i.e. the blue portion) to be the one given by subgroup $K_1$, and the boundary type to the right of $\mfL$ (red portion) to be given by $K_2$. In Section \ref{sec:hamiltonian-background}, we defined the Hamiltonian $H^{K}_{G}$ for each subgroup $K \subseteq G$, and demonstrate how to combine this with the original Kitaev Hamiltonian to produce the gapped boundary Hamiltonian $H_{\text{G.B.}}$ for the lattice with arbitrary holes/boundary types. Following that model, we will now combine several Hamiltonians $H^{K}_{G}$ and the original $H_{G}$ to form $H_{\text{dft}}$:

\begin{equation}
\label{eq:defect-hamiltonian-1}
H_{\text{dft}} = H_{G} (\mfB) + H^{K_1}_{G} (\mfr_1) + H^{K_2}_{G} (\mfr_2) + H^{K_1 \cap K_2}_{G} (\mfL).
\end{equation}

\noindent
Here, the region $\mathfrak{r}_1$ consists of all edges, vertices, and plaquettes within the blue shaded rectangle and along the blue border lines (but not along the purple $\mfL$). $\mathfrak{r}_2$ is defined similarly. The line $\mfL$ consists of all vertices and edges along the purple line that divides $\mfr_1$ from $\mfr_2$. As before, the bulk $\mfB$ will denote all other vertices, edges, and plaquettes (i.e. the black and white region). It is simple to check that the above Hamiltonian is also a commuting Hamiltonian.

With this construction, we say that the boundary defects live on the two cilia highlighted in orange in Fig. \ref{fig:defect-hamiltonian}\cite{Note2}, where a {\it cilium} (pl. {\it cilia}) is defined as a pair $(\mfv,\mfp)$ of a vertex and a plaquette to which it belongs.

\begin{figure}
\centering
\includegraphics[width = \textwidth]{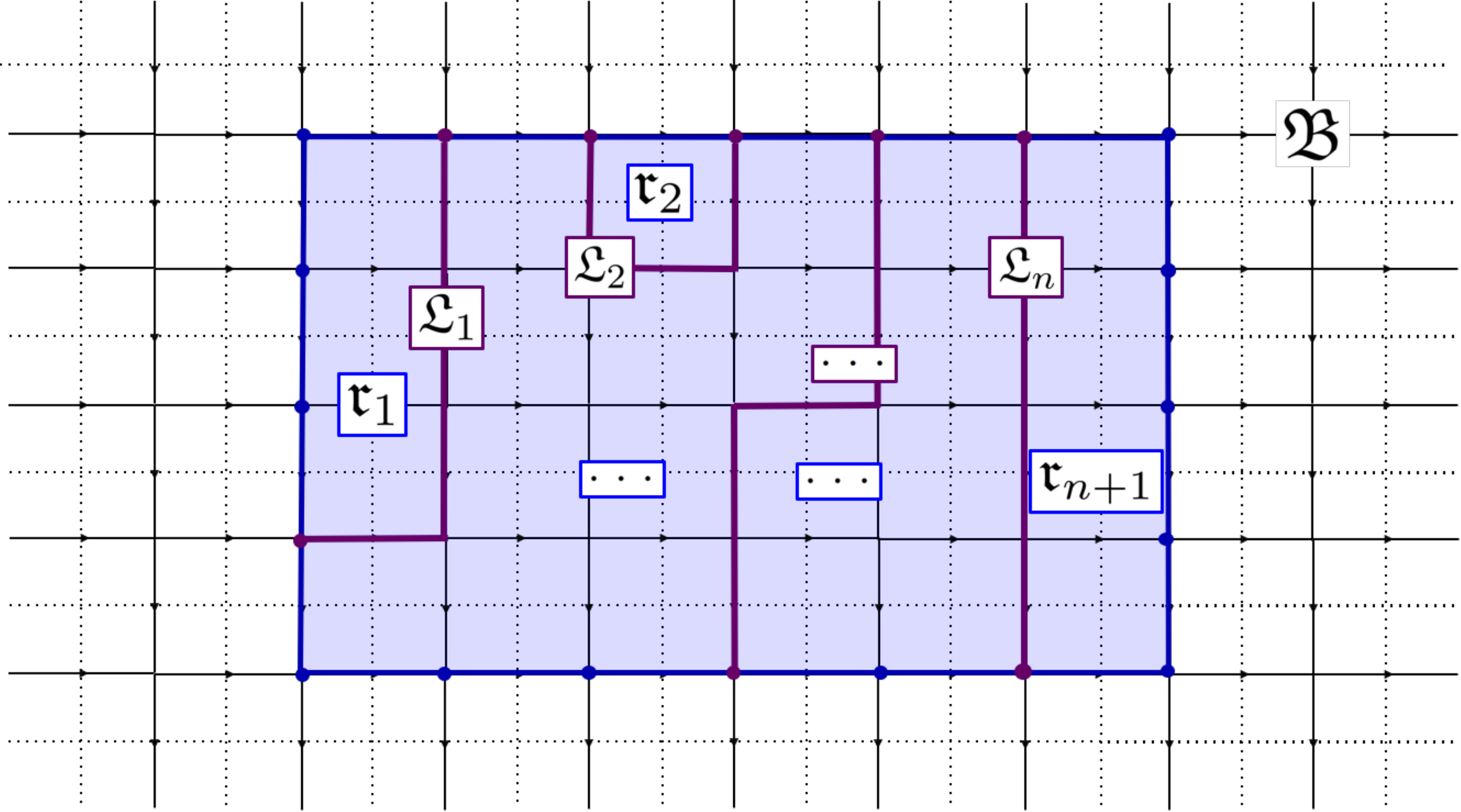}
\caption{More general definition of the Hamiltonian $H_{\text{dft}}$. There are $n$ non-intersecting piecewise-linear borders $\mfL_1, ... \mfL_n$, which separate the hole into $n+1$ regions $\mfr_1, ... \mfr_{n+1}$. Each $\mfL_i$ is formed from a chain of edges on the lattice or its dual. A total of $2n$ defects are created, one at each endpoint of each border $\mfL_i$. Each border $\mfL_i$ consists of all edges that the corresponding purple line(s) cross. Each region $\mathfrak{r}_j$ consists of all vertices, plaquettes, and edges within the corresponding blue shaded region, including the vertices and edges along the border of the hole, but not along the purple lines. The bulk $\mfB$ will denote all other vertices, edges, and plaquettes (i.e. the black and white region).} 
\label{fig:defect-hamiltonian-2}
\end{figure}

More generally, we may create as many defects as we like on the boundary of a hole, and the border lines between different boundary types need not be straight. This is illustrated in Fig. \ref{fig:defect-hamiltonian-2}. In this case, for each $i = 1, 2, ... n$, let us define $a(i), b(i) \in \{1,2,...n+1\}$ to be the numbers such that the regions on the two sides of $\mfL_i$ are $\mfr_{a(i)}$ and $\mfr_{b(i)}$ (it does not matter which is which). Suppose each region $\mfr_j$ is given by the subgroup $K_j \subseteq G$. The general Hamiltonian $H_{\text{dft}}$ is then defined as follows:

\begin{equation}
\label{eq:defect-hamiltonian-2}
H_{\text{dft}} = H_{G} (\mfB) + \sum_{j=1}^{n+1} H^{K_j}_{G} (\mfr_j) + \sum_{i = 1}^{n} H^{K_{a(i)} \cap K_{b(i)}}_{G} (\mfL_i).
\end{equation}

\noindent
Here, the Hamiltonian $H^{K_{a(i)} \cap K_{b(i)}}_{G}$ is applied to all vertices, plaquettes, and edges that the corresponding purple line(s) of $\mfL_i$ cross, including the vertex or plaquette on the boundary of the hole if applicable (purple dots in Fig. \ref{fig:defect-hamiltonian-2}). The Hamiltonian $H^{K_j}_{G}$ is applied to all vertices, plaquettes, and edges within the corresponding blue shaded region, including the edges along the border of the hole (blue in Fig. \ref{fig:defect-hamiltonian-2}), but not those along the purple lines. The bulk Hamiltonian is applied to all other vertices, edges, and plaquettes (i.e. the black and white region).

The generalization to the case with many holes is straightforward.

\begin{remark}
Note that the above Hamiltonian creates boundary defects in pairs (corresponding to the two endpoints of each purple line of Fig. \ref{fig:defect-hamiltonian-2}). In general, defects are quite similar to anyons, in the sense that they live on cilia (as opposed to gapped boundaries/holes). Hence, it seems impossible to create a single boundary defect, unlike gapped boundaries.
\end{remark}

\subsection{Topological properties of boundary defects}
\label{sec:simple-defect-topological}

We will now examine some topological properties of the defects between different boundary types. Let us consider again the simple case of two defects (the generalizations are obvious). As before, the two regions $\mfr_1$, $\mfr_2$ are given by subgroups $K_1, K_2 \subseteq G$, respectively. 

In Refs. \onlinecite{Cong16a,Cong16b}, we defined ribbon operators for the bulk and boundary Hamiltonians and used them to analyze topological properties of bulk and boundary excitations. In principle, one may do the same with the new Hamiltonians $H_{\text{dft}}$. The main purpose of this is to classify elementary excitations or simple defect types, which are defined as follows: Typically, a Hamiltonian like $H_{G}$ or $H_{\text{G.B.}}$ has a large space of possible excitations (similarly, $H_{\text{dft}}$ has a large space of possible defect types). An {\it elementary excitation type} (or {\it simple defect type}) is then one that cannot be changed under the action of local operators, and can only be changed by applying global {\it ribbon operators}. More details on this may be found in Sec. 2.2 of Ref. \onlinecite{Cong16a}.

For the case of $H_{\text{G.B.}}$, we performed a Fourier transform on the boundary ribbon operator algebra to show that the topological labels of the elementary excitations on a subgroup $K$ boundary are given by pairs $(T,R)$, where $T \in K \backslash G / K$ is a double coset, and $R$ is an irreducible representation of the stabilizer $K^{r_T} = r_T K r_T^{-1}$ for some representative $r_T \in T$. Similarly, in principle, one may perform a Fourier transform on the boundary defect ribbon operator algebra to obtain the corresponding simple defect types. The simple defect types are given by the set

\begin{equation}
\{
(T,R): T \in K_1 \backslash G / K_2, \text{ } R \in (K_1, K_2)^{r_T}_{\text{ir}} 
\}.
\end{equation}

\noindent
As before, $r_T \in T$ is a representative of the double coset. Here, we define the subgroup $(K_1, K_2)^{r_T} = K_1 \cap r_T K_2 r_T^{-1}$ to be the generalized stabilizer group. Furthermore, the quantum dimension of the simple defect is given by:

\begin{equation}
\label{eq:simple-defect-qdim}
\text{FPdim}(T,R) = \frac{\sqrt{|K_1| |K_2|}}{|K_1 \cap r_T K_2 r_T^{-1}|} \cdot \Dim(R).
\end{equation}

A proof of this formula can be found in Ref. \onlinecite{Yamagami}.  Note that in the case where $K_1 = K_2 = K$, the simple defect types are the same as the elementary excitations on the boundary.

\begin{remark}
In this analysis, we only considered defect types created using our Hamiltonian $H_{\text{dft}}$ for holes whose borders lie on the lattice. In the case of a corner defect (e.g. $\mfh_2$ in Fig. \ref{fig:boundary-hamiltonian-3}), one should first determine the subgroups that would give the corresponding boundary types if all boundary lines were on the lattice and then apply these formulas.
\end{remark}

\vspace{2mm}
\section{Algebraic Theory of Boundary Defects}
\label{sec:algebraic-theory}

In this section, we present an algebraic model for the defects between different gapped boundary types. Throughout the section, we will assume the reader is familiar with the concepts of fusion categories, modular tensor categories, and indecomposable module categories. We refer the reader to Refs. \onlinecite{Etingof15} and \onlinecite{BakalovKirillov} for these basic concepts.

\subsection{Gapped boundaries}
\label{sec:topological-order-gapped-boundaries}

Kitaev's quantum double model based on a finite group $G$ provides a commuting Hamiltonian to realize a topological phase of matter with topological order given by the modular tensor category $\B = \mfD(G) = \text{Rep}(D(G)) = \mZ(\text{Vec}_G) = \mZ(\text{Rep}(G))$ \cite{Kitaev97}. Alternatively, the Levin-Wen Hamiltonian \cite{Levin04} based on an input unitary fusion category $\mC$ gives rise to the topological order $\B = \mZ(\mC)$.

In Ref. \onlinecite{KitaevKong}, starting from the Levin-Wen model with input fusion category $\mC$, a method is presented to construct a gapped boundary for each indecomposable module category $\M$ of $\mC$. The converse, that every gapped boundary of the Levin-Wen model arises from an indecomposable module, is also conjectured there. For a gapped boundary given by the indecomposable module $\M$, the excitations of the boundary are given by objects in the functor category $\Fun_\mC(\M,\M)$, and the elementary excitations are the simple objects in this functor category. In this paper, we categorically describe the defects between two different gapped boundary types given by indecomposable modules $\M_i$ and $\M_j$.

As discussed in Sec. \ref{sec:hamiltonian-realization}, we must define an orientation for each gapped boundary, and without loss of generality, we define the orientation so that the bulk is always on the left hand side when we traverse the boundary. This makes it so that the gapped boundary will always be given by an indecomposable {\it left} module category of $\mC$.

\subsection{Boundary defects and the multi-fusion category}
\label{sec:multi-fusion-category}

It is  proposed in Ref. \onlinecite{KitaevKong} that in the Levin-Wen model based on input fusion category $\mC$, the defect types between two gapped boundaries given by indecomposable modules $\M_i$ and $\M_j$ are given by objects in the functor category $\mC_{ij} = \Fun_\mC(\M_i, \M_j)$.

The simple boundary defect types in a Kitaev model based on group $G$ between two boundaries given by subgroups $K_1, K_2 \subseteq G$ are given by pairs $(T,R)$, where $T \in K_1 \backslash G / K_2$ is a double coset, and $R$ is an irreducible representation of the stabilizer subgroup $(K_1, K_2)^{r_T} = K_1 \cap r_T K_2 r_T^{-1}$. By Ref. \onlinecite{Ostrik02}, if $\mC = \Rep(G)$, $\B = \mZ(\mC)$, and $\M_1$, $\M_2$ are indecomposable modules of $\mC$ corresponding by subgroups $K_1$ and $K_2$, respectively (with trivial cocycles), the simple objects of the bimodule category $\mC_{12} = \Fun_\mC(\M_1, \M_2)$ are given by pairs $(T,R)$, where $T \in K_1 \backslash G / K_2$ is a double coset, and $R$ is an irreducible representation of the subgroup $(K_1, K_2)^{r_T} = K_1 \cap r_T K_2 r_T^{-1}$. Because of this, we see that the boundary defects between these two boundary types are exactly given by the objects of the functor category $\mC_{12}$ in the Dijkgraaf-Witten theory.

In general, the category $\mC_{ij}$ is not a fusion category, as there is no canonical way to define a tensor product and dual within itself. However, if we consider all such functor categories over the input fusion category $\mC$, we get a $n \times n$ multi-fusion category $\mfC$, where $n$ is the number of inequivalent indecomposable modules. Such a unitary multi-fusion category has well-defined quantum dimensions for all simple objects; in the Dijkgraaf-Witten case, they are given by formulas of Section \ref{sec:hamiltonian-realization}.

\subsection{Topological degeneracy}
\label{sec:topological-degeneracy}

\begin{figure}
\centering
\includegraphics[width = 0.7\textwidth]{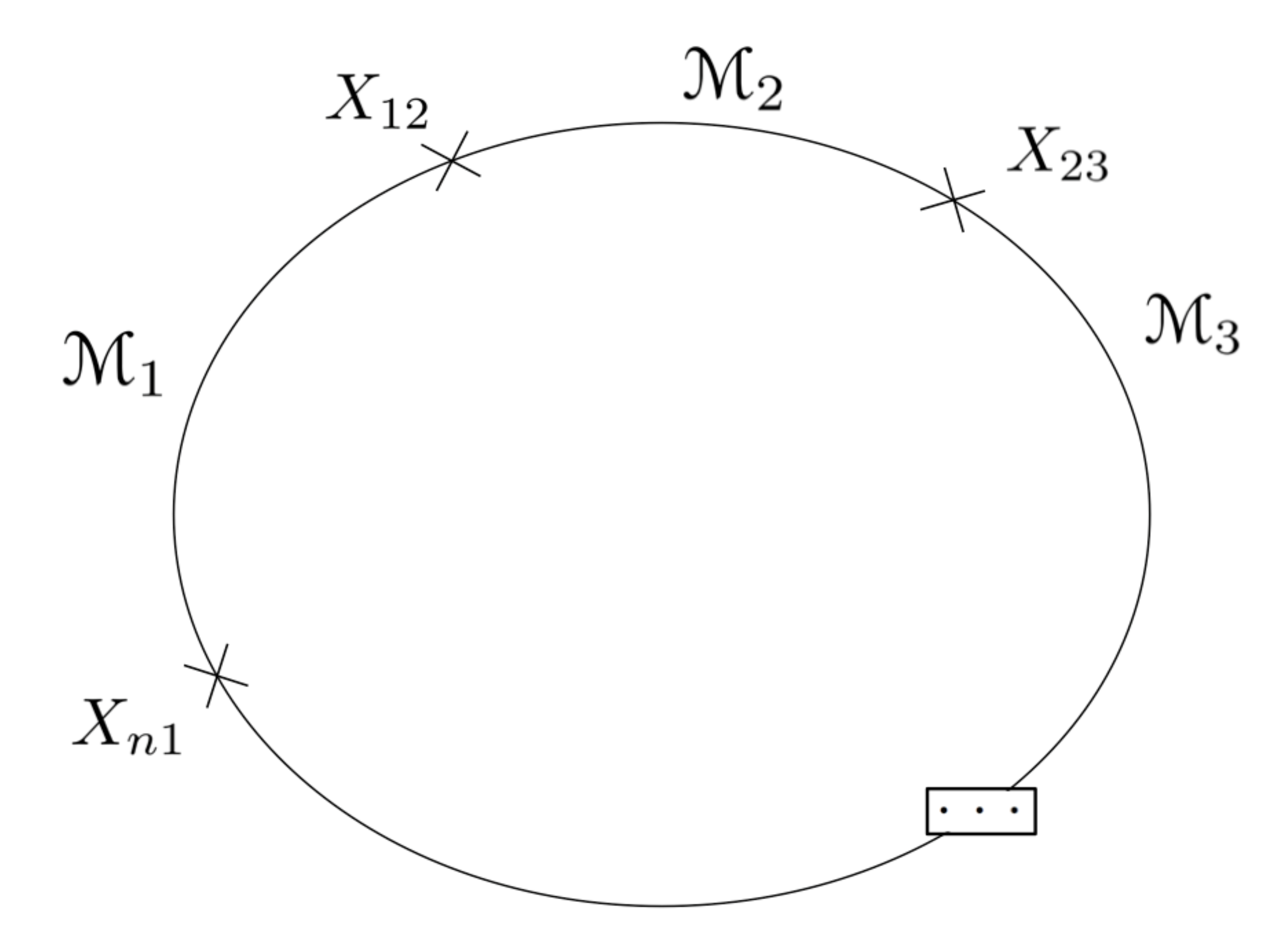}
\caption{Ground state degeneracy of $n$ boundary defects on a hole with $n$ boundary types given by indecomposable modules $\M_1$, ... $\M_n$.}
\label{fig:defects-fusion-2}
\end{figure}

As in the case of bulk anyons, topological degeneracy can also arise from the fusion of boundary defects. The most interesting case is when we have $n$ defects $X_{12} \in \mC_{12}$, $X_{23} \in \mC_{23}$, ... $X_{n1} \in \mC_{n1}$, and we would like to fuse them so that no boundary defect or excitation remains, and the resulting boundary is in the ground state of one of the boundaries (say the $\M_1$ boundary). The setup is illustrated in Fig. \ref{fig:defects-fusion-2}. Then, the topological degeneracy is given by the hom-space

\begin{equation}
\Hom(\one_{\M_1}, X_{12} \otimes X_{23} \otimes ... \otimes X_{n1})
\end{equation}

\noindent
where $\one_{\M_1}$ is the tensor unit in the fusion category $\mC_{11}$ (i.e. the trivial boundary excitation). This topological degeneracy may be used as a quantum memory or qudit encoding for the purposes of topological quantum computation (e.g. Ref. \onlinecite{Brown16}).

\subsection{Fusion of boundary defects}
\label{sec:fusion}

Let us suppose we have two boundary defects: $X_{12}$ between indecomposable modules $\M_1$ and $\M_2$, and $X_{23}$ between $\M_2$ and $\M_3$ (e.g. Fig. \ref{fig:defects-fusion-2}). As discussed in the previous sections, $X_{12} \in \mC_{12} = \text{Fun}_\mC(\M_1, \M_2)$, and $X_{23} \in \mC_{23} = \text{Fun}_\mC(\M_2, \M_3)$. Then, the fusion rules in the multi-fusion category $\mfC = \{\mC_{ij}\}$ tell us that fusing $X_{12}$ and $X_{23}$ gives the new defect

\begin{equation}
X_{13} = X_{12} \circ X_{23} \in \mC_{13} = \text{Fun}_\mC(\M_1, \M_3)
\end{equation}

\noindent
which is the composition of the original two defects. In general, of course, the result may not be a simple object in $\mC_{13}$, and we may see the decomposition $X_{13} = \oplus_{x_{13} \in \mC_{13}} n_{x_{13}} x_{13}$, where the coefficients $n_{x_{13}}$ would be given by the fusion rules of $\mfC$.

\vspace{2mm}
\section{Symmetry Defects and the Bulk-Edge Correspondence}
\label{sec:bulk-edge-correspondence}

With this algebraic theory of boundary defects, we now present a bulk-edge correspondence between symmetry defects in the bulk \cite{Barkeshli14} and defects between gapped boundaries, namely {\it crossed condensation}. This generalizes the notion of bulk-to-boundary condensation as a tensor functor \cite{Kong15,Cong16a,Cong16b}, and the nomenclature follows the definition of $G$-crossed braiding of symmetry defects in Ref. \onlinecite{Barkeshli14}.

\subsection{Gapped boundaries as Lagrangian algebras}
\label{sec:lagrangian-algebras}

In Section \ref{sec:topological-order-gapped-boundaries}, we stated that gapped boundary types in the topological order $\B = \mZ(\mC)$ can be given by indecomposable modules $\M$ of $\mC$. In examining bulk-edge correspondences, it is useful to take an alternative and equivalent view of gapped boundaries as Lagrangian algebras $\A$ in $\B$, which are defined as follows:

\begin{definition}
\label{lagrangian-algebra-def}
A object $\A$ with a multiplication $m: \A \otimes \A \rightarrow \A$ in a modular tensor category $\B$ is a {\it condensible algebra} if:
\begin{enumerate}
\item
$\A$ is {\it commutative}, i.e. $\A \otimes \A \xrightarrow{c_{\A\A}} \A \otimes \A \xrightarrow{m} \A$ equals $\A \otimes \A \xrightarrow{m} \A$, where $c_{\A\A}$ is the braiding in the modular category $\B$.
\item
$\A$ is {\it separable}, i.e. the multiplication morphism $m$ admits a splitting $\mu:\A \rightarrow \A \otimes \A$ a morphism of $(\A,\A)$-bimodules.
\item
$\A$ is {\it connected}, i.e. $\Hom_\B(\one_\B, \A) = \C$, where $\one_\B$ is the tensor unit of $\B$.
\end{enumerate}
A condensible algebra is {\it Lagrangian} if it is also a maximal such algebra, i.e.: if
\begin{enumerate}
\setcounter{enumi}{3}
\item
The Frobenius-Perron dimension of $\A$ is the square root of the dimension of the modular tensor category $\B$,
\begin{equation}
\label{eq:lagrangian-algebra-dim}
\FPdim(\A)^2 = \Dim(\B).
\end{equation}
\end{enumerate}
In the mathematical literature, condensible algebras are often known as special symmetric Frobenius algebras.
\end{definition}

By Proposition 4.8 of Ref. \onlinecite{Davydov12}, there exists a one-to-one correspondence between the indecomposable modules of $\mC$ and the Lagrangian algebras of $\B$, so it follows that gapped boundaries may equivalently be determined by Lagrangian algebras. Physically, because only bulk bosonic anyons in $\B$ may appear in the decomposition $\A = \oplus_a n_a a$ of $\A$ into simple objects, the gapped boundary may be viewed as a collection of these bulk bosonic anyons which can condense to vacuum on the boundary \onlinecite{Cong16a,Cong16b}. 

\subsection{Condensation as a tensor functor}
\label{sec:condensation-tensor-functor}

In Refs. \onlinecite{Cong16a,Cong16b}, we used a tensor functor, namely a quotient procedure, to describe the bulk-to-boundary condensation of anyons to boundary excitations and give the bulk-edge correspondence between these two types of excitations. Before we generalize that discussion to boundary defects, let us review the definition of this quotient functor \cite{MacLane}:

\begin{definition}
\label{quotient-cat-def}
Let $\mS$ be a tensor category, and let $\A$ be any object in $\mS$. The {\it pre-quotient category} $\widetilde{\mQ} = \mS/\A$ is the category such that:
\begin{enumerate}
\item
The objects of $\widetilde{\mQ}$ are the same as the objects of $\B$.
\item
The morphisms of $\widetilde{\mQ}$ are given by
\begin{equation}
\label{eq:quotient-hom-space}
\Hom_{\mS/\A}(X,Y) = \Hom_{\mS}(X,\A \otimes Y).
\end{equation}
\end{enumerate}
\end{definition}

We would like to describe the result of condensing bulk anyons in $\B$ to the gapped boundary $\A$ using the category $\B/\A$. However, one minor problem with this pre-quotient category is that the resulting category $\widetilde{\mathcal{Q}}$ may not be semisimple. The following definition/proposition of Ref. \onlinecite{Muger03} is hence needed to fully describe the condensation products:

\begin{definition}
\label{IC-def}
Let $\mS$ be a tensor category and let $\A$ be an object in $\mS$. Let $\widetilde{\mQ} = \mS/\A$ be the pre-quotient category formed via Definition \ref{quotient-cat-def}. Let us form the canonical idempotent completion ${\mQ}$ of $\widetilde{\mQ}$ as follows:
\begin{enumerate}
\item
The objects of ${\mQ}$ are given by pairs $(X,p)$, where $X \in \Obj \widetilde{\mQ}$ and $p = p^2 \in \End_{\widetilde{\mQ}}(X)$.
\item
The morphisms of ${\mQ}$ are given by
\begin{equation}
\begin{split}
\Hom_{{\mQ}}((X,p),(Y,q)) = \{ f \in \Hom_{\widetilde{\mQ}}(X,Y):\\
f \circ p = q \circ f\}. 
\end{split}
\end{equation}
\end{enumerate}
Then, by Proposition 2.15 of Ref. \onlinecite{Muger03}, the new category ${\mQ}$ is semisimple and the desired quotient.
\end{definition}

In Refs. \onlinecite{Cong16a,Cong16b}, condensation to a gapped boundary given by the Lagrangian algebra $\A$ is mathematically described as the procedure

\begin{equation}
\label{eq:condensation-quotient-IC}
F: \mZ(\mC) = \B \xrightarrow{\text{quotient}} \B/\A = \widetilde{\mQ} \xrightarrow{\text{I.C.}}  {\mQ}.
\end{equation}

\noindent
(Here, I.C. is the idempotent completion). The proof that the quotient category $\mQ$ is indeed the excitation category $\text{Fun}_\mC(\M,\M)$ (where $\M$ is the indecomposable module corresponding to $\A$) is given in Ref. \onlinecite{Davydov12}. Furthermore, the right adjoint $I$ of this procedure, which is given by the composition of the adjoint of the idempotent completion and the adjoint $\widetilde{I}$ of the quotient functor, is exactly the inverse condensation procedure where a boundary excitation leaves the boundary and enters the bulk.

\subsection{Symmetry defects in the bulk}
\label{sec:symmetry-defects}

Symmetry defects in the bulk of topological phases were studied extensively in Ref. \onlinecite{Barkeshli14}; we briefly review the material that is relevant to our discussion here.

Starting from a braided tensor category $\B$, a braided tensor auto-equivalence $\phi: \B \rightarrow \B$ is an invertible monoidal functor that permutes the simple objects $\phi(a) = a'$ while preserving all topological properties (e.g. quantum dimension and twist). A global symmetry group $G$\cite{Note3}
of $\B$ is then a group with an equivalence class of homomorphisms $[\rho]: G \rightarrow \bta (\B)$ from $G$ to the group of braided tensor auto-equivalences of $\B$, where the equivalence relation is under natural transformations of $\B$. Given a braided tensor category $\B$, the Picard group $\pic(\B)$ is the group of invertible (bi-)module\cite{Note4}
categories over $\B$. By Ref. \onlinecite{Etingof10}, there is an isomorphism

\begin{equation}
\label{eq:bta-brpic}
\bta(\B) \cong \pic(\B).
\end{equation}

Given a global symmetry $G$ of $\B = \mZ(\mC)$, if all obstructions vanish, one can introduce symmetry defects into the bulk anyon theory \cite{Barkeshli14}. This enriches the topological phase so that it is now described by a $G$-graded fusion category $\B_G = \oplus_{g \in G} \B_g$, where $\B_0 = \B$ is the original modular tensor category, whose simple objects are the original bulk anyons. By the isomorphism (\ref{eq:bta-brpic}), each $\B_g$ is an invertible bi-module category in $\pic(\B)$ corresponding to the braided tensor auto-equivalence $\rho_g \in \bta(\B)$, and simple objects ($g$-defects) in $\B_g$ are the symmetry defects which carry a ``flux'' $g$. The fusion and associativity of these defects respect the multiplication of $G$, so that a $g$-defect and $h$-defect fuse to a $gh$-defect; specifically, the $G$-graded fusion \cite{Barkeshli14} of two defects is given by the tensoring of bi-modules over $\B$. Finally, the (right-handed) braiding of defects in $\B_G$ is a $G$-crossed \cite{Barkeshli14} braiding\cite{Note5}:

\begin{equation}
\label{eq:G-crossed-braiding}
\vcenter{\hbox{\includegraphics[width = 0.38\textwidth]{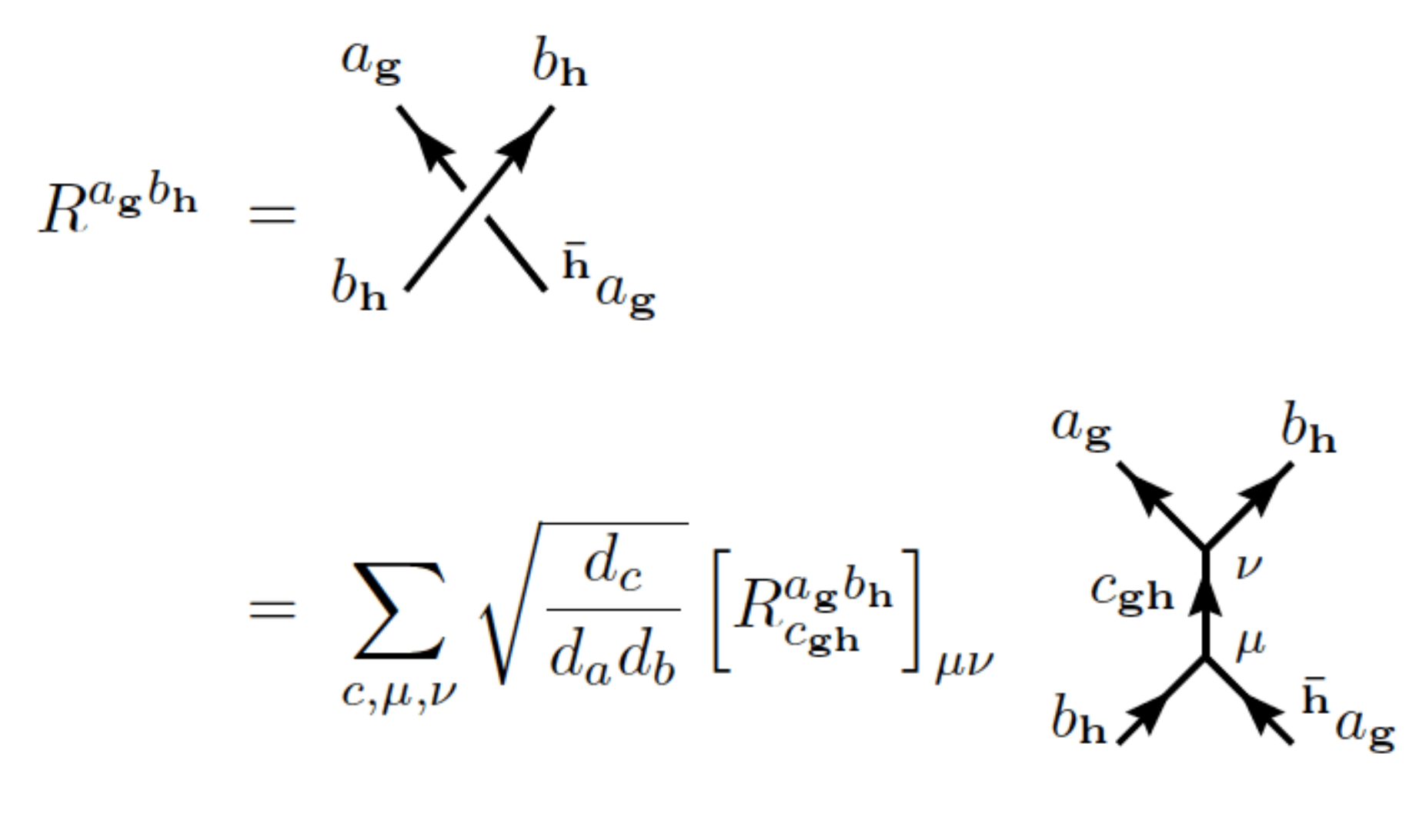}}}
\end{equation}

\noindent
In the above equation, following the notation of Ref. \onlinecite{Barkeshli14}, $a_g$ indicates that $a \in \B_g$, $\overbar{g} = g^{-1}$, $^g h = ghg^{-1}$, and $^g b_h = \rho_g (b_h)$. The $R$ symbols $R^{a_g b_h}_{c_{gh}}$ in the $G$-graded category are the maps for the counter-clockwise exchange of topological charges $b_h$ and $a_{^{\overbar{h}} g}$ from the fusion channel $b_h \otimes a_{^{\overbar{h}} g} \rightarrow c_{gh}$ to the fusion channel $a_g \otimes b_h \rightarrow c_{gh}$. In particular, note that the braiding results in a $h$-action on $a_{^{\overbar{h}} g}$, which potentially changes its topological label. A similar definition exists for the left-handed braiding.

\subsection{Crossed condensation}
\label{sec:crossed-condensation}

We now describe the bulk-edge correspondence between certain types of boundary defects and symmetry defects in the bulk.

Let $\M_i$, $\M_j$ be indecomposable module categories of the input fusion category $\mC$ (assume $i \neq j$). Suppose $\M_i$ and $\M_j$ are such that $\mC_{ii} = \text{Fun}_\mC(\M_i, \M_i)$ and $\mC_{jj} = \text{Fun}_\mC(\M_j, \M_j)$ are equivalent as fusion categories. The category $\mC_{ij} = \text{Fun}_\mC(\M_i, \M_j)$ is an invertible $\mC_{ii}-\mC_{jj}$ bi-module, and we assume that $\mC_{ij}$ is a nontrivial element in the Brauer-Picard group $\brpic(\mC_{ii})$ of invertible $\mC_{ii}-\mC_{ii}$ bi-modules.
Theorem 2.4 of Ref. \onlinecite{Chang15} tells us that $\mC$ and $\mC_{ii}$ are Morita equivalent. By Ref. \onlinecite{Etingof10}, we then have the isomorphisms

\begin{equation}
\brpic(\mC_{ii}) \cong \bta(\mZ(\mC_{ii})) = \bta(\mZ(\mC)).
\end{equation}

\noindent
It follows that the equivalence of $\mC_{ii}$ and $\mC_{jj}$ induces a $\Z_2$ symmetry in the bulk theory $\mZ(\mC_{ii}) = \mZ(\mC)$.

As discussed in the previous section, if all obstructions vanish, we can construct a $\Z_2$-graded fusion category $\B_{\Z_2} = \B_0 \oplus \B_1$ where $\B_0 = \B = \mZ(\mC)$ is the original category and $\B_1$ is the nontrivial (i.e. not $\B$) bi-module category in $\pic(\B)$. Using the definition of the quotient category discussed in Section \ref{sec:condensation-tensor-functor}, we can study what happens to a bulk symmetry defect when it is dragged to the gapped boundary given by indecomposable module $\M_i$. Specifically, let $\A_i$ be the corresponding Lagrangian algebra. Taking $\mS = \B_{\Z_2}$ in Definitions \ref{quotient-cat-def} and \ref{IC-def}, we adopt the procedure of Eq. (\ref{eq:condensation-quotient-IC}) and perform a quotient by $\A_i$ followed by the idempotent completion to obtain the semi-simple category $\mQ(\Z_2, \A_i)$ which describes the condensed objects.

By Eq. (\ref{eq:quotient-hom-space}), since $\A_i$ is an object in the trivial flux sector, only objects in the same flux sector of $\B_{\Z_2}$ can be identified with each other in this process; hence, we may write $\mQ(\Z_2, \A_i) = \mQ_0(\A_i) \oplus \mQ_1(\A_i)$ for the results of condensing the two sectors. Furthermore, since the objects in $\mQ_0(\A_i)$ are simply the results of condensing the objects in $\B_0$ (i.e. the bulk anyons), $\mQ_0(\A_i)$ is the category of boundary excitations on the gapped boundary $\A_i$, i.e. $\mQ_0(\A_i) = \mC_{ii}$. Similarly, the bulk symmetry defects arising from the $\Z_2$ symmetry $\mC_{ii} \cong \mC_{jj}$ condense to the boundary as boundary defects in $\mQ_1(\A_i) \subset \mC_{ij}$.  We conjecture $\mQ_1(\A_i) = \mC_{ij}$.

The generalization of Eq. (\ref{eq:condensation-quotient-IC}) to the {\it crossed condensation} procedure is hence

\begin{equation}
\begin{split}
\label{eq:crossed-condensation-quotient-IC}
F: \B_{\Z_2} \xrightarrow{\text{quotient}} \B_{\Z_2}/\A_i = \widetilde{\mQ}(\Z_2, \A_i)\\
\xrightarrow{\text{I.C.}}  {\mQ}(\Z_2, \A_i) = \mC_{ii} \oplus \mC_{ij}.
\end{split}
\end{equation}

Similarly, the right adjoint $I$ of $F$ describes the result of pulling a boundary excitation in $\mC_{ii}$ or a boundary defect in $\mC_{ij}$ to the bulk: it becomes a bulk anyon in $\B_0$ or a bulk symmetry defect in $\B_1$, respectively (see Fig. 5.1).

\floatsetup[figure]{style=plain,subcapbesideposition=bottom}
\begin{figure}
\label{fig:crossed-condensation}
     \centering
        \sidesubfloat[]{%
            \includegraphics[width=0.45\textwidth]{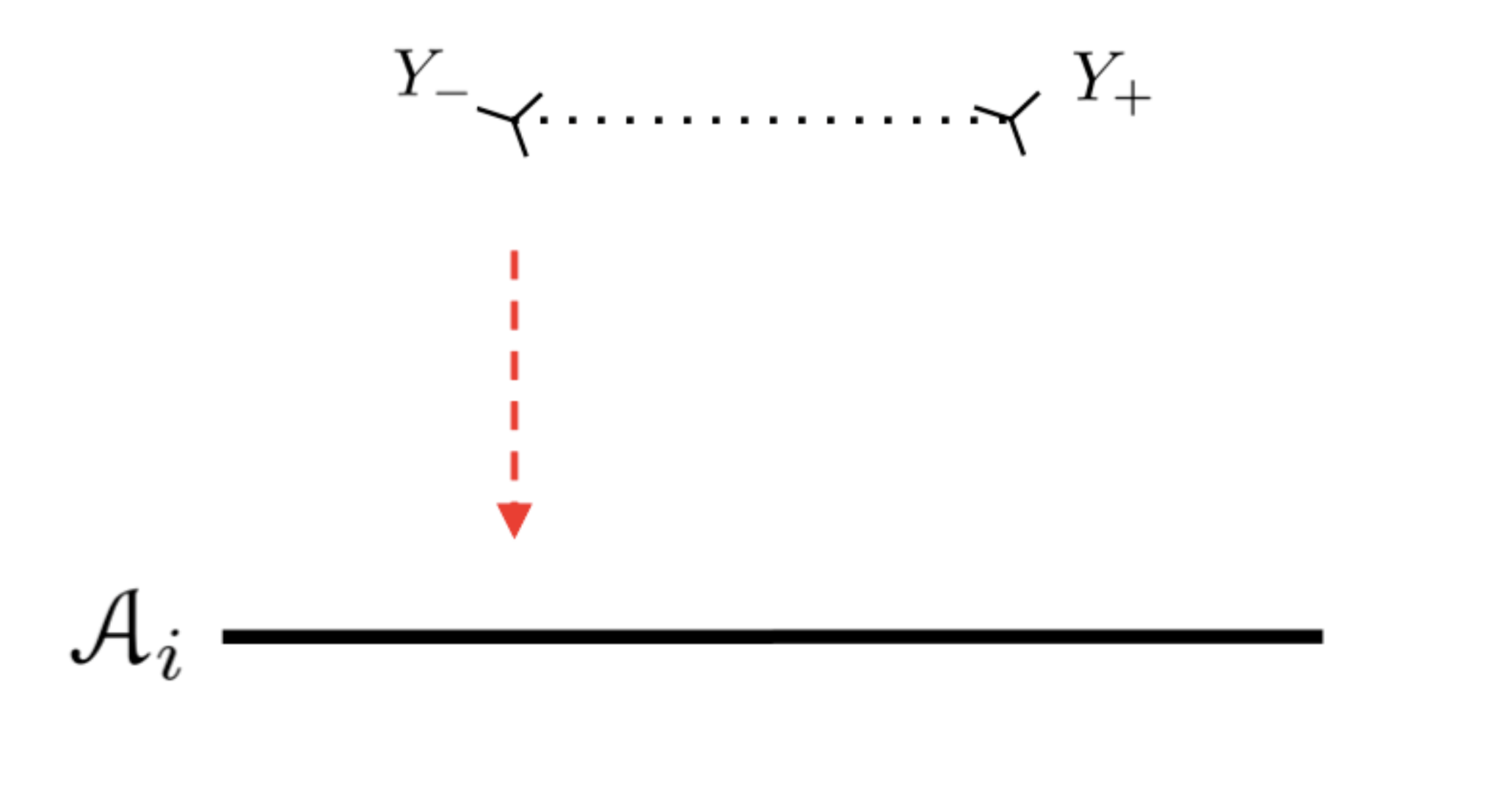}\label{fig:crossed-condensation-a}
        }%
        \sidesubfloat[]{%
           \includegraphics[width=0.45\textwidth]{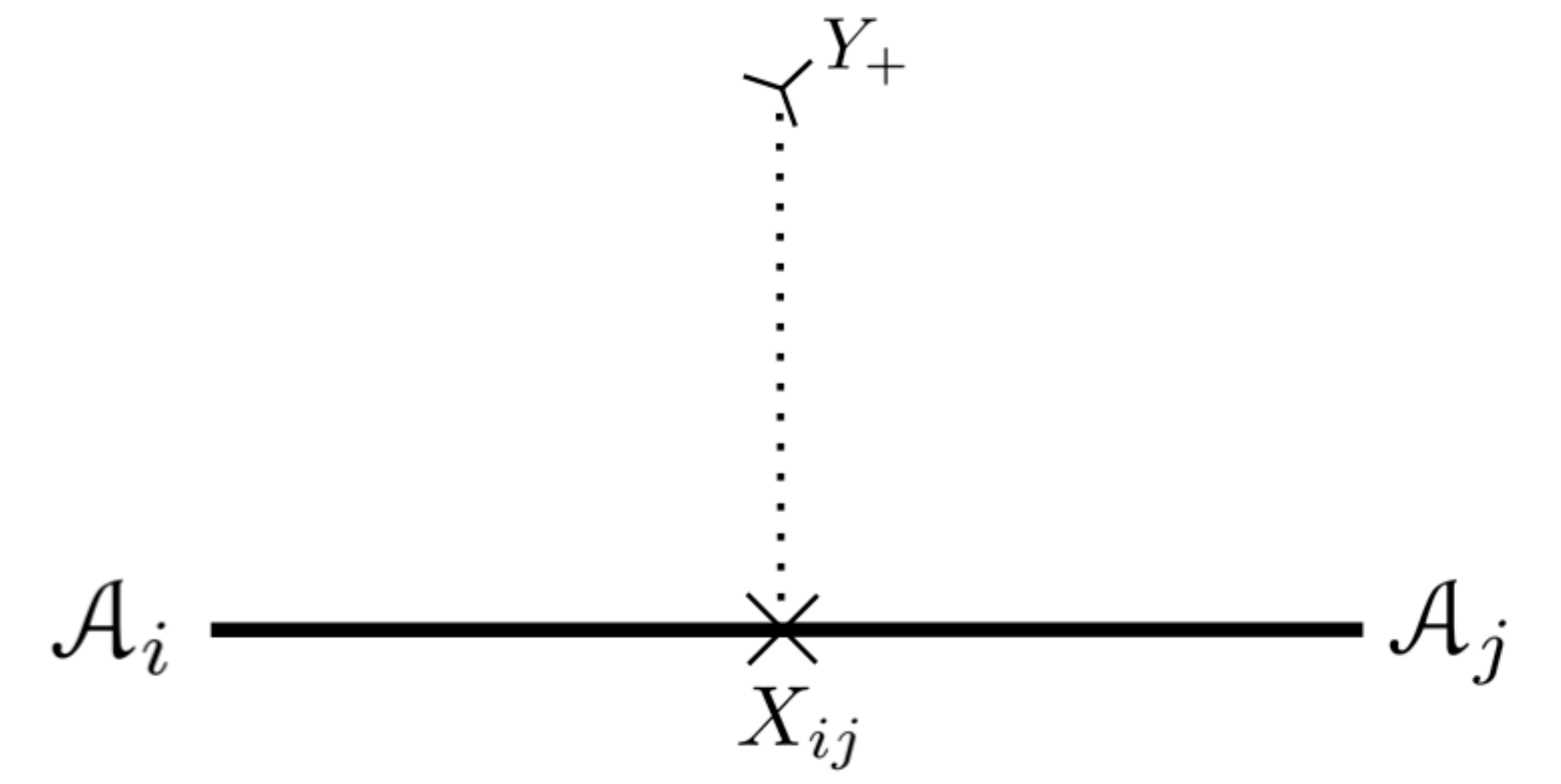}\label{fig:crossed-condensation-b}
        }\\ 
    \caption{Crossed condensation. (A) A symmetry defect $Y_-$ and its dual $Y_+$ in the bulk. (B) After the crossed condensation, the bulk symmetry defect turns in to a boundary defect. Furthermore, the condensation procedure modifies one side of the gapped boundary $\A_i$ by the $\Z_2$ action to produce the gapped boundary $\A_j$.}
\end{figure}

We have now described how bulk anyons correspond to boundary excitations, and how bulk symmetry defects correspond to boundary defects.  This leads to one interesting puzzle: how can a bulk symmetry defect condense to a boundary defect $X_{ij} \in \mC_{ij}$, when $X_{ij}$ must be located at a defect site between {\it two different} gapped boundary types $\A_i$ and $\A_j$, and we only have one gapped boundary type $\A_i$ in the picture (e.g. Fig. \ref{fig:crossed-condensation-a})? This can be explained perfectly by viewing the $\Z_2$-crossed braiding of symmetry defects (see Eq. (\ref{eq:G-crossed-braiding})) in the context of gapped boundaries.

In Section \ref{sec:lagrangian-algebras}, we mentioned that viewing the gapped boundary as a Lagrangian algebra $\A_i = \oplus_a n_a a$ describes the bulk bosonic anyons that can condense to vacuum on the boundary. As discussed in Section \ref{sec:symmetry-defects}, when a bulk anyon $a$ is exchanged counter-clockwise with a symmetry defect in the flux sector $\B_g$, $a$ undergoes a $g$-action (potentially changing to a different topological label $a' = \rho_g(a)$). The same occurs in our context: the condensation procedure pushes the bulk symmetry defect slightly below the boundary (i.e. to a cilium within the boundary, to use the terminology of Section \ref{sec:hamiltonian-realization}). Because of this, the gapped boundary line (or the Lagrangian algebra $\A_i$) undergoes a $\Z_2$ action, transforming one side of the boundary to the gapped boundary type $\A_j = \oplus_a n_a \rho_1(a)$, as in Fig. \ref{fig:crossed-condensation-b} ($\rho_1$ is the action by the non-identity $\Z_2$ element). Specifically, since the boundary is oriented left-to-right in Fig. \ref{fig:crossed-condensation-a} (Sec. \ref{sec:algebraic-theory}), by Eq. (\ref{eq:G-crossed-braiding}), the right hand side undergoes the action and is the one changed to $\A_j$.

Through this correspondence, we see how an object in the invertible bi-module category $\B_1$ of $\B = \mZ(\mC) = \mZ(\mC_{ii}) = \mZ(\mC_{jj})$ becomes an invertible $\mC_{ii}-\mC_{jj}$ bi-module when condensed to the boundary. We believe this is related to the fusion of domain walls, as the defect line connecting $Y_-$ to $Y_+$ in Fig. \ref{fig:crossed-condensation-a} is essentially a (nontrivial) domain wall between $\B$ and itself, and this condensation procedure describes the fusion of the domain wall with the gapped boundary $\A_i$ (i.e. a domain wall between $\B$ and $\text{Vec}$).

While we have focused here on the case of $\Z_2$ symmetries in the bulk, the generalization to arbitrary bulk symmetry groups $G$ is straightforward. In that case, we would start with the $G$-graded category $\B_G = \oplus_{g \in G} \B_g$, which describes the bulk anyon theory when enhanced with symmetry defects. In condensing a symmetry defect in the sector $\B_g$, the gapped boundary would undergo a $g$-action, modifying the right hand side of the boundary in Fig. \ref{fig:crossed-condensation-b} to $\A_{j_g} = \oplus_a n_a \rho_g(a)$. Each defect sector $\B_g$ in the bulk would be sent to the category $\mC_{ij_g} = \Fun_\mC(\M_i, \M_{j_g})$ of boundary defects, where $\M_{j_g}$ is the indecomposable module corresponding to the Lagrangian algebra $\A_{j_g}$. Because $\rho_g$ preserves the topological properties of objects in $\B$, the fusion categories $\mC_{j_g j_g} = \mQ(1, \A_{j_g})$ will be isomorphic to $\mC_{ii} = \mQ(1, \A_{i})$, and the boundary defect categories $\mC_{ij_g}$ will again be the invertible $\mC_{ii}-\mC_{ii}$ bi-module categories in the Brauer-Picard group $\brpic(\mC_{ii})$. The generalized crossed condensation functor would then be

\begin{equation}
\begin{split}
\label{eq:crossed-condensation-quotient-IC-G}
F: \B_{G} \xrightarrow{\text{quotient}} \B_{G}/\A_i = \widetilde{\mQ}(G, \A_i)\\
\xrightarrow{\text{I.C.}}  {\mQ}(G, \A_i) = \oplus_{g \in G} \mC_{i j_g}.
\end{split}
\end{equation}

\vspace{2mm}
\section{Braiding of Boundary Defects}
\label{sec:braiding}

In this section, we discuss the projective braiding of certain boundary defects using the algebraic theory of Section \ref{sec:algebraic-theory}. Our computation of the projective braiding uses the crossed condensation theory of Section \ref{sec:bulk-edge-correspondence} and the $G$-crossed braiding of Ref. \onlinecite{Barkeshli14}. We then discuss possible physical implementations of this braiding using forced measurements as in Ref. \onlinecite{Bonderson13} and their potential applications to topological quantum computation. We also show how bulk anyons may be braided with boundary defects.

\subsection{Assumptions and setup}

Let $\B = \mZ(\mC)$ be a bulk anyon system which is a quantum double/Drinfeld center. Suppose $\B$ has a global symmetry group $G$ as discussed in Section \ref{sec:symmetry-defects}, and suppose we have a gapped boundary given by Lagrangian algebra $\A_i$ (or equivalently, indecomposable module $\M_i$ of $\mC$). We can then form the $G$-graded fusion category $\B_G = \oplus_{g \in G} \B_g$.

Fix some $g \in G$; since $\A_i \in \B = \B_0 \subset \B_G$, we can form a second Lagrangian algebra 

\begin{equation}
\label{eq:gapped-bd-symmetry}
\A_j = \rho_g (\A_i)
\end{equation}

\noindent
which determines another gapped boundary type. In the following, we consider the braiding of defects in $\mC_{ij} = \Fun_\mC(\M_i, \M_j)$ with those in $\mC_{ji}$.

By Ref. \cite{Chang15}, we have the isomorphism

\[
\mZ(\mfC) \cong \mZ(\mC) = \B
\]

\noindent
where $\mfC$ is the multi-fusion category of all boundary excitations and defects. It follows that the global symmetry group $G$ of $\B$ also acts on the boundary defects. In particular, the next section discusses the braiding of boundary defects in $\mC_{ij}$ and $\mC_{ji}$ in separate cases, depending on whether the defects in $\mC_{ij}$ are fixed by the action of $g \in G$.

\subsection{Braiding amongst boundary defects: the main definition/theorem}
\label{sec:braiding-main-def-thm}

We now present a definition/theorem governing the braiding of boundary defects in the multi-fusion category $\mfC$. 

\begin{defthm}
\label{braid-def-thm}
Let $\A_i$ and $\A_j$ be two Lagrangian algebras (gapped boundaries) in $\B = \mZ(\mC)$, and let $\M_i$ and $\M_j$ be the corresponding indecomposable module categories. Suppose $\A_i$ and $\A_j$ are related by a global $G$ symmetry of $\B$, as in Eq. \ref{eq:gapped-bd-symmetry}. 
Then there is a projective $G$-crossed braiding of the boundary defects in $\mC_{ij}$ with those in $\mC_{ji}$, and with the boundary excitations in $\mC_{jj}$. Furthermore, there is a canonical choice of this braiding and a systematic method to compute the projective representation, discussed in Section \ref{sec:compute-braiding}.\\
In particular, if all defects in $\mC_{ji}$ are fixed by the action of $g \in G$ as discussed in the previous section, the projective $G$-crossed braiding is simply a projective braiding of boundary defects.
\end{defthm}

The proof of this theorem will be given in the next section, where we provide a systematic computation of the projective braiding. We first present a straightforward yet important corollary:

\begin{corollary}
Let $\M_i$, $\M_j$ be indecomposable module categories of the input fusion category $\mC$. Suppose in addition that
$\mC_{ii} = \text{Fun}_\mC(\M_i, \M_i)$ and $\mC_{jj} = \text{Fun}_\mC(\M_j, \M_j)$ are equivalent as fusion categories, and $\mC_{ij} \neq \mC_{ii}$. As discussed in Section \ref{sec:crossed-condensation}, there is then a $\Z_2$ symmetry in the bulk $\B$. If all boundary defects in $\mC_{ji}$ are fixed by the $\Z_2$ action, the boundary defects in $\mC_{ij}$ can be projectively braided with those in $\mC_{ji}$, and with the boundary excitations in $\mC_{jj}$. Furthermore, there is a canonical choice of this braiding and a systematic method to compute the projective representation, discussed in Section \ref{sec:compute-braiding}.
\end{corollary}

\subsection{Computing the projective braiding}
\label{sec:compute-braiding}

Because neither the multi-fusion category $\mfC$ nor its components $\mC_{ii}$, $\mC_{ij}$ are modular categories, there is no obvious way to braid boundary defects by looking simply at these categories.
However, in Section \ref{sec:crossed-condensation}, we developed a correspondence between boundary defects $X_{ij} \in \mC_{ij}$ and bulk $G$ symmetry defects . Since we know the $G$-graded category $\B_G = \oplus_{h \in G} B_h$ has a canonical $G$-crossed braiding for the bulk, we solve the boundary defects braiding problem by mapping the defects into the bulk, and using this $G$-crossed braiding.

In Section \ref{sec:crossed-condensation}, we provided a functor $F$ which described the condensation of a bulk object in $\B_{G}$ to a boundary excitation or boundary defect, and the adjoint $I$ of $F$ described the ``inverse condensation'' of a boundary excitation or defect into the bulk. Suppose $X_{ij} \in \mC_{ij}$, $X_{ji} \in \mC_{ji}$. Then, the canonical $G$-crossed braiding of these two defects posited in the main definition/theorem is given by 

\begin{equation}
\begin{gathered}
\label{eq:boundary-defects-braiding}
X_{ij} \otimes X_{ji} \xrightarrow{I \otimes I} I(X_{ij}) \otimes I(X_{ji}) \\
\xrightarrow{G^\times \text{braid}} \rho_g(I(X_{ji})) \otimes I(X_{ij})
\xrightarrow{F \otimes F} \rho_g(X_{ji}) \otimes X_{ij}.
\end{gathered}
\end{equation}

\noindent
Specifically, the boundary defects are first brought into the bulk to $I(X_{ij}), I(X_{ji}) \in \B_g$, then braided in $\B_{G} = \oplus_h \B_h$ via the $G$-crossed braiding ($\rho_g$ is again the action of $g \in G$), and finally condensed back to $\rho_g(X_{ji})$, $X_{ij}$ on the boundary. The projective $G$-crossed braiding of the two defects is computed precisely as the $G$-crossed braiding of their images in the bulk $\B_{G}$.

The same computation holds when the boundary defect $X_{ji}$ is replaced by a boundary excitation $X_{jj} \in \mC_{jj}$, so Eq. \ref{eq:boundary-defects-braiding} also provides the braiding of $X_{ij}$ with $X_{jj}$ posited in the main definition/theorem.

Finally, it is clear that if all elements in $\mC_{ji}$ are fixed under the action of $g \in G$, then $\rho_g (X_{ji}) = X_{ji}$, so this projective $G$-crossed braiding of the two defects is simply a projective braiding.

\subsection{Physical implementations and applications}
\label{sec:physical-implementations-applications}

In the previous sections, we have developed a mathematical description of the braiding of boundary defects under the conditions of Definition/Theorem \ref{braid-def-thm}. However, since we do not know an immediate and suitable way to physically implement this braiding or the crossed condensation procedure, we now discuss a potential mechanism to simulate the braiding using forced measurements and ancilla boundary defects. In the special case of parafermion zero modes, Ref. \onlinecite{lindner2012} has presented a similar realization in a fractional quantum Hall-superconductor setup.

\begin{figure}
\centering
\includegraphics[width = 0.75\textwidth]{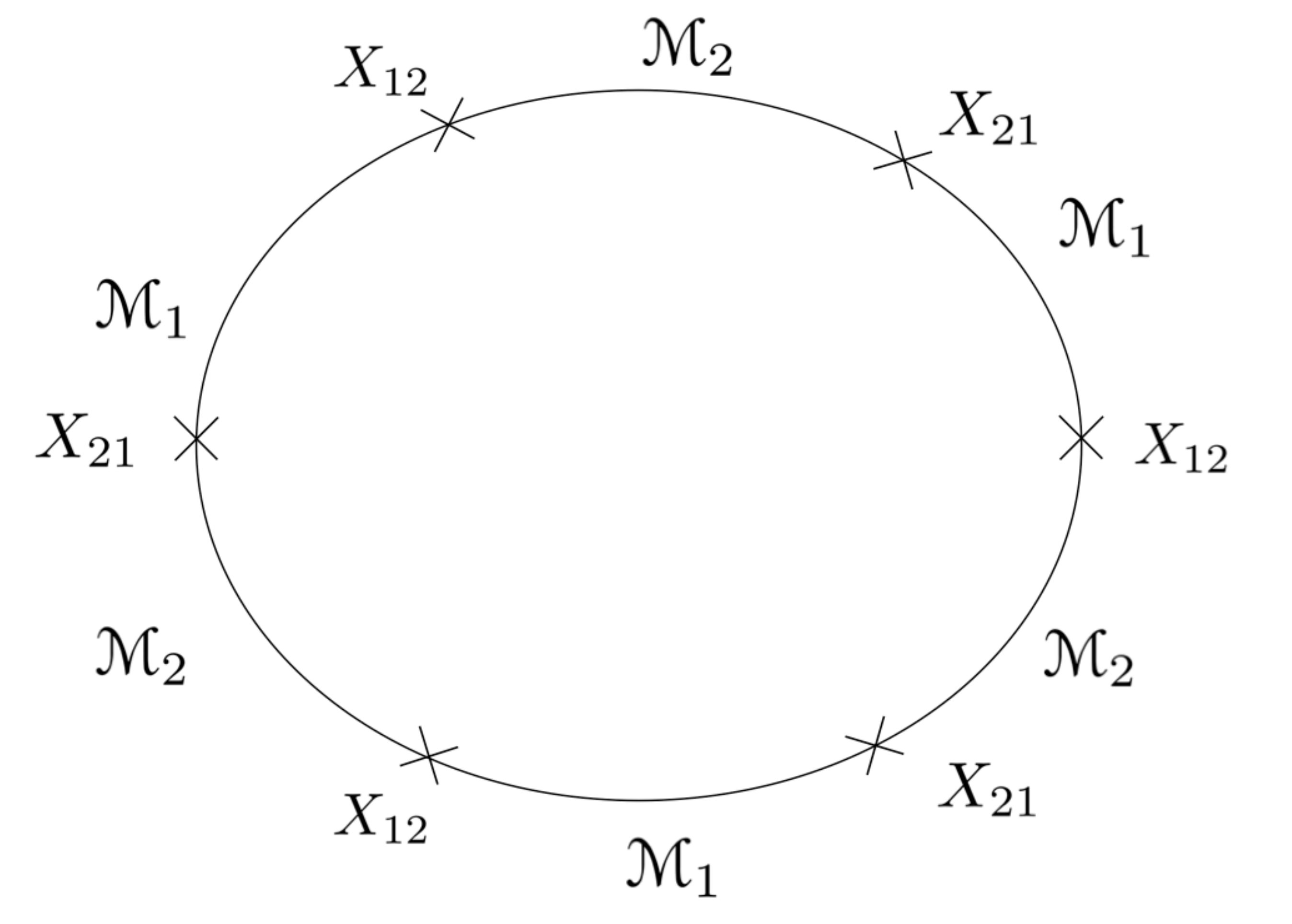}
\caption{Six boundary defects on a ring. The mathematical braiding of two defects $X_{12}$ and $X_{21}$ may be implemented physically by introducing ancilla boundary defects and using forced measurements.}
\label{fig:physical-braiding}
\end{figure}

Let us start with six boundary defects on the boundary of a disk, as shown in Fig. \ref{fig:physical-braiding}. The fusion of boundary defects was discussed in Section \ref{sec:fusion} as composition of functors in the multi-fusion category $\mfC$. By this fusion rule, the result of fusing $X_{12} \otimes X_{21}$ to a defect $X_{11} = \oplus_{x_{11} \in \mC_{11}} n_{x_{11}} x_{11}$ in $\mC_{11}$ need not be a simple object; because of this, we assume that we can perform ``forced measurements'' which physically project to a specific fusion channel $X_{12} \otimes X_{21} \rightarrow x_{11}$. With this assumption, Ref. \onlinecite{Bonderson13} shows how one may use three $X_{12}$ boundary defects and one $X_{21}$ boundary defect (i.e. two $X_{12}$ ancillas) to simulate the braiding of $X_{12}$ with $X_{21}$.

In Section \ref{sec:topological-degeneracy}, we discussed how the topological degeneracy of boundary defects could be used to encode a topological quantum memory for the purposes of topological quantum computation. In that setting, the above physical implementation of boundary defect braiding could provide topological quantum gates. Such topological operations may be of significant interest in specific cases such as the boundary defect realization of Majorana zero modes or genons (see Section \ref{sec:examples}).

\subsection{Braiding bulk anyons with boundary defects}
\label{sec:braid-anyons-defects}

Thus far, we have discussed how to braid certain boundary defects with each other. In fact, if we assume the defect category $\mC_{ij}$ satisfies the conditions of Definition/Theorem \ref{braid-def-thm}, it is also possible to braid bulk anyons with boundary defects. Specifically, the braiding of the bulk anyon $a$ with a boundary defect $X_{ij} \in \mC_{ij}$ again uses the crossed condensation theory:

\begin{equation}
\begin{gathered}
a \otimes X_{ij} \xrightarrow{\Id \otimes I} a \otimes I(X_{ij})\\
\xrightarrow{G^\times \text{braid}} I(X_{ij}) \otimes \rho_1(a)
\xrightarrow{F \otimes \Id} X_{ij} \otimes \rho_1(a).
\end{gathered}
\end{equation}

As before, we pull the defect out of the boundary to $I(X_{ij}) \in \B_1$, and then use the $\Z_2$-crossed braiding of $\B_{\Z_2} = \B_0 \oplus \B_1$ to braid it with $a \in \B_0$, and finally condense $I(X_{ij})$ back to the boundary. Note that the bulk anyon $a$ will undergo the $\Z_2$ action in this procedure, and may be permuted to a different topological label.

\vspace{2mm}
\section{Important Examples}
\label{sec:examples}

\subsection{Majorana and Parafermion Zero Modes Revisited}
\label{sec:majorana-parafermion}

In Section \ref{sec:majorana-parafermion-hamiltonian}, we presented a local commuting projector Hamiltonian to realize Majorana/parafermion zero modes as corners of gapped boundaries in the $\Z_p$ toric code. We now revisit this example by using the theory of Sections \ref{sec:algebraic-theory}-\ref{sec:braiding} to investigate the algebraic/topological properties of these boundary defects and further illustrate their equivalence with Majorana/parafermion zero modes.

\subsubsection{Algebraic theory}
\label{sec:parafermion-algebraic-theory}

\begin{table}
\caption{Multi-fusion category $\mfC$ for the input category $\mC = \text{Vec}_{\Z_p}$. The diagonal entries are functor fusion categories $\mC_{ii} = \Fun_\mC(\M_i, \M_i)$. The off-diagonal entries are abelian semisimple categories $\mC_{ij} = \Fun_\mC(\M_i, \M_j)$; for these categories, we provide the quantum dimensions of the simple objects in curly braces.}\label{tab:dzp-multi-fusion}
\centering
\begin{tabular}{|c|c|c|}
\hline
   $\boldsymbol{\mfC}$ & $\boldsymbol{\A_1}$ & $\boldsymbol{\A_2}$ \\ \hline
   $\boldsymbol{\A_1 = 1+e+...+e^{p-1}}$ / $\boldsymbol{K_1 = \{1\}}$ & $\text{Vec}_{\Z_p}$ & $\left\{\sqrt{p}\right\}$ \\ \hline
   $\boldsymbol{\A_2 = 1+m+...+m^{p-1}}$ / $\boldsymbol{K_2 = \Z_p}$ & $\left\{\sqrt{p}\right\}$ & $\text{Rep}(\Z_p)$ \\ \hline
\end{tabular}
\end{table}

Let us first construct the multi-fusion category $\mfC = \{ \mC_{ij} = \Fun_\mC(\M_i,\M_j) \}$ that contains all the boundary excitations and defects. 

The topological order of the $\Z_p$ toric code is given by the modular tensor category $\mfD(\Z_p) = \mZ(\text{Vec}(\Z_p)) = \mZ(\Rep(\Z_p))$, where the simple objects are of form $e^i m^j$ ($i,j = 0,1, ... p-1$), and the fusion rules are $e^{i_1} m^{j_1} \otimes e^{i_2} m^{j_2} \rightarrow e^{i_1+i_2} m^{j_1+j_2}$ (addition is done modulo $p$).
As discussed in Section \ref{sec:majorana-parafermion-hamiltonian}, the $\Z_p$ toric code has only two gapped boundary types corresponding to subgroups $K_1 = \{0\}$, $K_2 = \Z_p$; Refs. \onlinecite{Cong16a,Cong16b} show that these correspond to Lagrangian algebras $\A_1 = 1+e+...+e^{p-1}$ and $\A_2 = 1+m+...+m^{p-1}$, respectively. Let $\M_i$ be the indecomposable module category over $\text{Vec}_{\Z_p}$ corresponding to $\A_i$. Table \ref{tab:dzp-multi-fusion} presents the quantum dimensions of the simple objects in each component of $\mfC$.

\subsubsection{The bulk-edge correspondence}
\label{sec:parafermion-bulk-edge}

We now examine the bulk-edge correspondence between boundary defects and symmetry defects in the $\Z_p$ toric code. As seen in Table \ref{tab:dzp-multi-fusion}, the two boundary excitation fusion categories are isomorphic: $\mC_{11} = \text{Vec}(\Z_p) \cong \Rep(\Z_p) = \mC_{22}$, so we hope to find a correspondence between boundary defects in $\mC_{12}$ and $\Z_2$ symmetry defects in $\mfD(\Z_p)$. Let ${\tau_{ij}}$ denote the one simple object of $\mC_{ij}$ (i.e. the simple defect where $\A_i$ is on the left of $\A_j$); let $1,X,...,X^{p-1}$ denote the simple objects (i.e. boundary excitations) in $\mC_{11}$ with fusion rules $X^i \otimes X^j \rightarrow X^{i+j(\text{mod}p)}$.

By Ref. \onlinecite{Barkeshli14}, we can enhance $\B = \mfD(\Z_p)$ by introducing a symmetry defect sector $\B_1$ corresponding to the electric-magnetic $\Z_2$ symmetry. There are $p$ simple objects $\tau_0$, $\tau_1$, ... $\tau_{p-1}$ in $\B_1$, each of quantum dimension $\sqrt{p}$, which differ by fusion of an $e$ or $m$ anyon. The condensation of bulk anyons/symmetry defects to the boundary $\A_i$ is then given by:

\begin{equation}
e^i m^j \rightarrow X^j, \qquad \tau_{i} \rightarrow \tau_{12}
\end{equation}

Similarly, the adjoint of the condensation (i.e. boundary excitations/defects leaving the boundary) is given by:

\begin{equation}
m^j \rightarrow \oplus_{i = 0}^{p-1} e^i m^j, \qquad
\tau_{12} \rightarrow \oplus_{i=0}^{p-1} \tau_i.
\end{equation}

The same computation may be done for the gapped boundary $\A_2$ and the boundary defects $\tau_{21} \in \mC_{21}$; the result is the same with $e,m$ switched.

\subsubsection{Braiding}
\label{sec:parafermion-braiding}

Using the above bulk-edge correspondences, we can now apply the method of Section \ref{sec:braiding} to compute the projective braiding of the simple defect $\tau_{12} \in \mC_{12}$ with the simple defect $\tau_{12} \in \mC_{21}$. In the case of $p = 2$, the $R$ symbols for the $\Z_2$-crossed braiding in $\B_{\Z_2}$ \cite{Barkeshli14} tell us that this projective braiding gives a $\pi/16$ phase, which is also consistent with the braid statistics of the Majorana zero mode.

\subsubsection{Exact degeneracy}
\label{sec:parafermion-exact-degeneracy}

While the asymptotic degeneracy of many boundary defects on the boundary of a disk is determined by the quantum dimension of the defect, the exact degeneracy has to be obtained differently. Table \ref{dz2-exact-degeneracy} shows the exact degeneracy of $2n$ boundary defects on the edge of a disk, for the case of $\mfD(\Z_2)$. Below, we explain how this degeneracy may be obtained by counting fusion channels.

\begin{table}\caption{Ground state degeneracy of $2n$ defects on the edge of a disk, between the $1+e$ and $1+m$ boundaries of the $\Z_2$ toric code.}\label{dz2-exact-degeneracy}
\centering
\begin{tabular}{|c|c|c|}
\hline
   $\boldsymbol{\mfC}$ & $\boldsymbol{\A_1}$ & $\boldsymbol{\A_2}$ \\ \hline
   $\boldsymbol{\A_1}$ & $1$ & $2^{n-1}$  \\ \hline
   $\boldsymbol{\A_2}$  & $2^{n-1}$ & $1$ \\ \hline
\end{tabular}
\end{table}

Excitations on a gapped boundary $\A_i$ form a fusion category $\mC_{ii}$. The vacuum is the condensate, while excitations are like solitons. Notice that the fusion of boundary excitations are not necessarily commutative. Mathematically, the boundary fusion category $\mathcal{C}_{ii}$ is the representation category $\text{Rep}(\mathcal{A}_i)$ of the Lagrangian algebra $\mathcal{A}_i$ that characterizes the condensate. By Ref. \onlinecite{Chang15}, $\mathcal{Z}(\mathcal{C}_{ii})$ must recover the bulk topological order.

Consider two different Lagrangian algebras $\mathcal{A}_i, \mathcal{A}_j$, and denote a boundary defect between $\mathcal{A}_i$ and $\mathcal{A}_j$ as $\tau_{ij}$. For each $\mathcal{A}_i$, we can find the boundary fusion category $\text{Rep}(\mathcal{A}_i)$ with simple objects $\{t_i\}$ , and the fusion coefficients $\{n_a^{t_i}\}$:
\begin{equation}
	a\rightarrow \sum_i n_a^{t_i}t_i.
\end{equation}
\noindent
These coefficients can be systematically computed for Dijkgraaf-Witten theories.

The fusion of $\tau_{ij}$ with $\tau_{ji}$ physically shrinks the $\mathcal{A}_j$ segment (sandwiched between two $\mathcal{A}_i$ segments) to a point. The result should be an excitation of the $\mathcal{A}_i$ boundary, so we can write
\begin{equation}
\label{eq:defect-fusion-coeffs}
	\tau_{ij}\otimes\tau_{ji}=\sum_{t_i}N_{\tau_{ij} \tau_{ji}}^{t_i} t_i.
\end{equation}
\noindent
Similarly, $\tau_{ji}\otimes\tau_{ij}$ can be defined. 

\begin{figure}
	\centering
	\includegraphics[width=0.5\columnwidth]{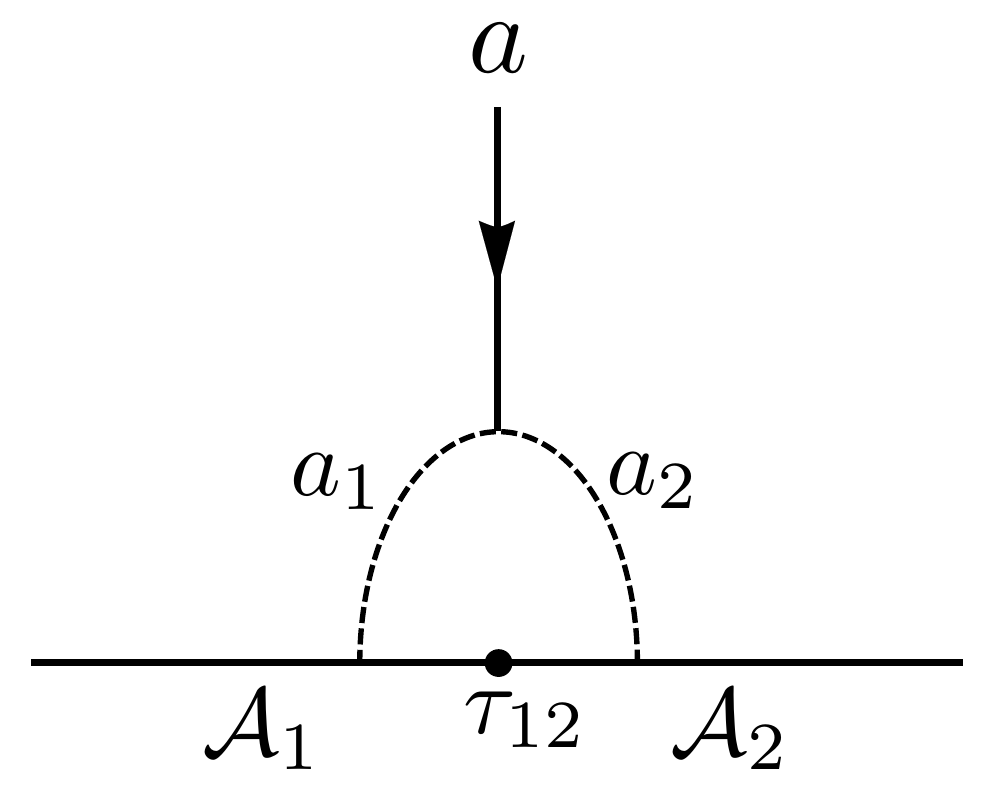}
	\caption{Illustration of a defect site absorbing a bulk anyon $a$.}
	\label{fig:zeromode}
\end{figure}

We now provide a heuristic method to determine the coefficients $N_{\tau_{ij} \tau_{ji}}^{t_i}$. We first find the ``zero modes'' of the boundary defects. Zero modes are bulk anyons which can be fused into the boundary defect. Namely, if $a\in\mathcal{A}_1\otimes\mathcal{A}_2$ (i.e. there exists $a_1\in \mathcal{A}_1$ and $a_2\in\mathcal{A}_2$ such that $a\in a_1\otimes a_2$), $a$ is a zero mode of $\tau_{12}$ (and $\tau_{21}$). The absorption process of $a$ by a defect site $\tau_{12}$ is illustrated in Fig. \ref{fig:zeromode}.  We conjecture that

\begin{conj}
\label{zero-mode-conj}
The simple object $t_i$ appears in $\tau_{ij}\otimes\tau_{ji}$ if and only if all $a$ in the lifting of $t_i$ (i.e. $n_a^{t_i}\neq 0$) are zero modes of $\tau_{ij}$ (or $\tau_{ji}$).
\end{conj}

The $\Z_2$ toric code $\mfD(\Z_2)$ has four anyon types $1,e,m,\psi=e\otimes m$. Condensation to the boundary $\A_1 = 1+e$ is given by 

\begin{equation}
	1/e\mapsto (1+e) \qquad m/\psi\mapsto (m+\psi).
	\label{}
\end{equation}

\noindent
where $(1+e)$ is the vacuum, and $(m+\psi) = X$ is the single simple boundary excitation. Similar results hold for condensation to the boundary $\A_2 = 1+m$.

On the interface between an $e$-boundary and an $m$-boundary, the defect site can absorb a $\psi$ particle. Fusion of $\tau_{12}$ and $\tau_{21}$ then amounts to shrinking the $m$-segment inside an $e$-segment, and the result is an excitation in the $e$-segment:

\begin{equation}
	\tau_{12}\otimes\tau_{21}=1+X.
	\label{}
\end{equation}

We may then use this fusion rule to compute the exact degeneracy of $2n$ boundary defects on a disk, by counting the number of fusion trees with total charge $1$. We denote the number of fusion trees with a total charge $a$ by $D_n^a$. We get the recursion relation

\begin{equation}
	\begin{pmatrix}
		D_n^1\\
		D_n^X
	\end{pmatrix}=
	\begin{pmatrix}
		1 & 1 \\
		1 & 1
	\end{pmatrix}
\begin{pmatrix}
	D_{n-1}^1\\
	D_{n-1}^X
	\end{pmatrix}.
	\label{}
\end{equation}

\noindent
Since $D^1_1 = D^X_1 = 1$, we have

\begin{equation}
	\begin{pmatrix}
		D_n^1\\
		D_n^X
	\end{pmatrix}=
	\begin{pmatrix}
		1 & 1 \\
		1 & 1
	\end{pmatrix}^{n-1}
\begin{pmatrix}
	1\\
	1
	\end{pmatrix}.
	\label{}
\end{equation}

\noindent
which gives $D^1_n = D^X_n = 2^{n-1}$.

\subsection{Genons}
\label{sec:genons}

Another important application of boundary defects is its equivalence to genons in a bilayer TQFT. As discussed in Ref. \onlinecite{Bark13a}, a genon in a bilayer TQFT is a defect in the $\Z_2$ symmetry that interchanges the two layers of the theory. Genons have been studied extensively in Refs. \onlinecite{Barkeshli11, Bark13a}, and can allow for universal quantum computation (see Refs. \onlinecite{Bark13a, Barkeshli15}). However, these groups have almost exclusively constructed genons for bilayer abelian group TQFTs, and did not provide an exactly solvable Hamiltonian realization. In this section, we extend these works in two ways: We first present a commuting projector Hamiltonian to realize genons in $\mfD(G \times G)$ as boundary defects, where $G$ is a finite abelian group. We then generalize the discussion of genons to bilayer TQFTs starting from an arbitrary input unitary fusion category $\mC$.

\begin{figure}
\centering
\includegraphics[width = 0.9\textwidth]{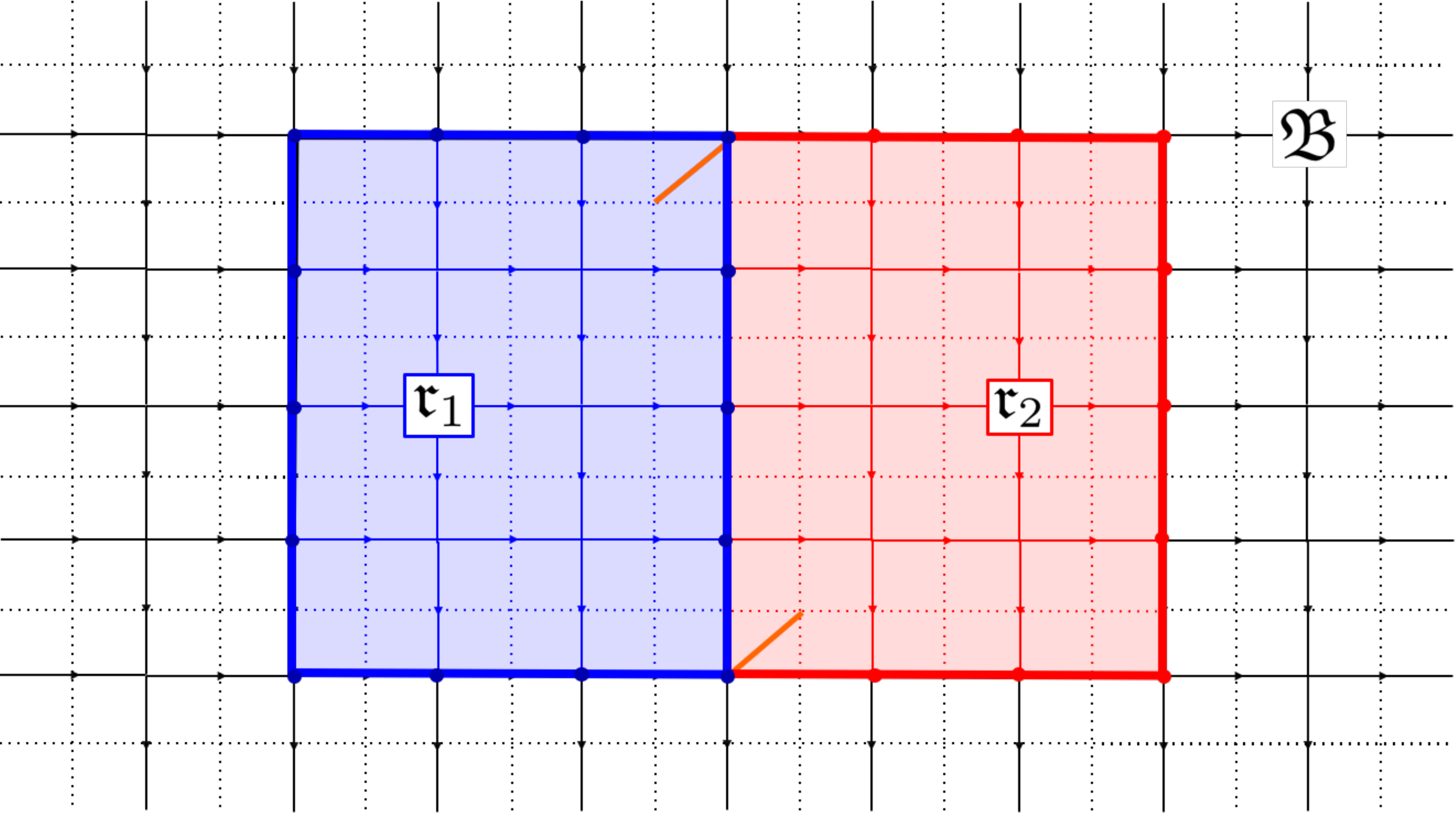}
\caption{Definition of the Hamiltonian $H_{\text{gn}}$. The region $\mfr_1$ (resp. $\mfr_2$) consists of all blue (red) vertices, plaquettes, and edges. The bulk $\mfB$ consists of everything else (black and white). The genons are created on the two orange cilia.} 
\label{fig:genon-hamiltonian}
\end{figure}

\subsubsection{Hamiltonian realization in Dijkgraaf-Witten theories}
\label{sec:genon-hamiltonian}
 
The Hamiltonian to create boundary defect genons is a special case of the general defect Hamiltonian $H_{\text{dft}}$, in the case where the input group is $G \times G$ (i.e. all data qudits on edges take values in the Hilbert space $\C[G \times G]$) for some finite abelian group $G$. In this situation, we begin with a hole in the lattice, divided into two regions, $\mfr_1$ and $\mfr_2$, as pictured in Fig. \ref{fig:genon-hamiltonian}. The region $\mfr_1$ consists of all vertices and plaquettes within and on all boundaries of the blue shaded rectangle, and all blue edges, including the ones on the thick blue dividing line between the two regions. The vertices and plaquettes of the region $\mfr_2$ are all those within the red shaded rectangle and on the upper, right, or lower boundaries of the rectangle; the edges of $\mfr_2$ are all red edges. To create genons, we consider two specific subgroups of $G \times G$, namely the trivial subgroup $K_1 = \{1_G\} \times \{1_G\}$ and the diagonal subgroup $K_2 = \{ (g,g): g \in G \}$ ($1_G$ is the identity element of $G$). We apply the Hamiltonian $H^{K_1}_{G \times G}$ to the region $\mfr_1$, and the Hamiltonian $H^{K_2}_{G \times G}$ to the region $\mfr_2$; as always, the bulk Hamiltonian $H_{G \times G}$ is applied to the bulk $\mfB$. Hence, the Hamiltonian to produce two boundary defect genons is given by 

\begin{equation}
\label{eq:genon-hamiltonian}
H_{\text{gn}} = H_{G \times G} (\mfB) + H^{K_1}_{G \times G} (\mfr_1) + H^{K_2}_{G \times G} (\mfr_2).
\end{equation}

\noindent
Of course, the generalization to producing multiple boundary defect genons on the same hole is straightforward.

By Section \ref{sec:hamiltonian-realization}, the simple defect types that can be created by $H_{\text{gn}}$ are given by pairs $(T,R)$, where $T \in K_1 \backslash G / K_2$ is a double coset, and $R$ is an irreducible representation of $(K_1, K_2)^{r_T} = K_1 \cap r_T K_2 r_T^{-1}$ for some representative $r_T \in T$. Since $K_1$ is trivial, $(K_1, K_2)^{r_T}$ is always trivial, and there are exactly $|G|$ double cosets $T$, one corresponding to each $r_T = (1_G,g)$, $g \in G$. By Eq. (\ref{eq:simple-defect-qdim}), the quantum dimension for each simple boundary defect genon is

\begin{equation}
\label{eq:genon-qdim}
\Dim(T,R) = \frac{\sqrt{|K_1| |K_2|}}{|K_1 \cap r_T K_2 r_T^{-1}|} \cdot \Dim(R) = \sqrt{|G|}
\end{equation} 

\noindent
This agrees precisely with the prediction of Section X.H in Ref. \onlinecite{Barkeshli14}, which states (for group-theoretical cases) that there should be exactly $|G|$ defects. In particular, the boundary defect genon given by $r_T = (1_G,1_G)$ and the trivial representation corresponds precisely with the ``bare defect'' described in Ref. \onlinecite{Barkeshli14}; it is consistent with the result that bare defect genon is a direct sum of all simple objects in the modular tensor category formed by giving $\C[G]$ a braided structure (and hence has quantum dimension $\sqrt{|G|}$).

\subsubsection{Algebraic theory: the general case}
\label{sec:genon-algebraic}

Let us now consider any system with topological order $\B$ given by the Drinfeld center of a bilayer $\D$ of a modular tensor category $\mC$. This can be done as follows: Since $\mE = \mC \boxtimes \overbar{\mC}$ is already a modular tensor category, let us choose a Lagrangian algebra $\A$ of $\mE$. We now construct two gapped boundaries (Lagrangian algebras) $\A_{1234}$, $\A_{1423}$ in the MTC $\B$: 

\begin{equation}
\begin{gathered}
\A_{1324} = \A_{13} \boxtimes \A_{24} \\
\A_{1423} = \A_{14} \boxtimes \A_{23}
\end{gathered}
\end{equation}

\noindent
Here, $\A_{ij}$ is the Lagrangian algebra corresponding to $\A$ when considered in the modular tensor category $\mE_{ij} = \mC_i \boxtimes \overbar{\mC_j} \cong \mE$. As indecomposable module categories of $\B$, we have

\begin{equation}
\begin{gathered}
\M_{1324} = \M_{13} \boxtimes \M_{24} \\
\M_{1423} = \M_{14} \boxtimes \M_{23}
\end{gathered}
\end{equation}

\noindent
where $\M_{ij}$ is the indecomposable module category in $\mE_{ij}$ corresponding to the Lagrangian algebra $\A_{ij}$. To generalize the language used by Ref. \onlinecite{Bark13a}, $\A_{1324}$ represents the {\it intralayer} gapped boundary, while $\A_{1423}$ represents the {\it interlayer} gapped boundary. Then, the boundary defect genon corresponds to a simple defect between these two gapped boundaries, i.e. it is a simple object in the functor category $\Fun_\D(\M_{1324}, \M_{1423})$. 

\subsubsection{The bulk-edge correspondence}

We mentioned earlier that the genons in Refs. \onlinecite{Barkeshli11, Bark13a} were defects in $\Z_2$ symmetry of $\mfD(G \times G)$ that interchanged the two layers of the bilayer TQFT. In our realizations, however, our ``boundary defect genons'' have been boundary defects between two gapped boundaries in $\mfD(G \times G)$. While our realization may seem different from the traditional definition, by Section \ref{sec:crossed-condensation}, the boundary defect genons are related to the $\Z_2$ symmetry defect genons simply by the condensation procedure $F$ and its adjoint $I$ (see Eq. (\ref{eq:crossed-condensation-quotient-IC})).

Genons play a very important role in quantum computation, as their braiding, when combined with anyons, has the power to give universal quantum computation (see Ref. \onlinecite{Barkeshli16}). By Section \ref{sec:braiding}, our boundary defect genons would yield the same braiding statistics as the $\Z_2$ symmetry defect genons and can also be braided with bulk anyons. Hence, this realization may also be of great importance to topological quantum computation.

\subsection{$\mfD(S_3)$}
\label{sec:ds3}

As another example, let us consider the simplest non-abelian group-theoretical case, namely the Dijkgraaf-Witten theory with $G = S_3$. In this case, the topological order of the resulting Kitaev model is given by the modular tensor category $\B = \mfD(S_3) = \Rep(D(S_3)) = \mZ(\Rep(S_3))$. As discussed in Ref. \onlinecite{Cui15}, there are 8 simple objects in this category, $A,B,...,H$. The fusion rules \cite{Cui15} are given in Table \ref{tab:DS3-fusion}.

{\tiny
\begin{table*}\caption{Fusion rules of $\mfD(S_3)$}\label{tab:DS3-fusion}
\begin{tabular}{|c|c|c|c|c|c|c|c|c|}
\hline $\boldsymbol{\otimes}$ &$\boldsymbol{A}$ &$\boldsymbol{B}$ &$\boldsymbol{C}$ &$\boldsymbol{D}$ &$\boldsymbol{E}$ &$\boldsymbol{F}$ &$\boldsymbol{G}$ &$\boldsymbol{H}$\\ \hline
$\boldsymbol{A}$ &$A$ &$B$ &$C$ &$D$ &$E$& $F$ &$G$ &$H$\\ \hline
$\boldsymbol{B}$ &$B$ &$A$ &$C$& $E$ &$D$ &$F$ &$G$ &$H$\\ \hline
$\boldsymbol{C}$ &$C$ &$C$ &$A\oplus B\oplus C$& $D\oplus E$ &$D\oplus E$ & $G\oplus H$& $F\oplus H$ &$F\oplus G$\\ \hline
\multirow{2}{*}{$\boldsymbol{D}$} &\multirow{2}{*}{$D$} &\multirow{2}{*}{$E$} &\multirow{2}{*}{$D\oplus E$}& $A\oplus C\oplus  F$ & $B\oplus C\oplus F$ & \multirow{2}{*}{$D\oplus E$} & \multirow{2}{*}{$D\oplus E$} & \multirow{2}{*}{$D\oplus E$} \\
& & & & $\oplus G\oplus H$ & $\oplus G\oplus H$ & & &  \\ \hline
\multirow{2}{*}{$\boldsymbol{E}$} &\multirow{2}{*}{$E$}& \multirow{2}{*}{$D$}& \multirow{2}{*}{$D\oplus E$} & $B\oplus C\oplus F$ & $A\oplus C\oplus F$ & \multirow{2}{*}{$D\oplus E$} &\multirow{2}{*}{$D\oplus E$} & \multirow{2}{*}{$D\oplus E$} \\
& & & & $\oplus G\oplus H$ & $\oplus G\oplus H$ & & &  \\  \hline
$\boldsymbol{F}$ &$F$ & $F$& $G\oplus H$& $D\oplus E$ & $D\oplus E$ & $A\oplus B\oplus F$ & $H\oplus C$ & $G\oplus C$ \\ \hline
$\boldsymbol{G}$ &$G$ & $G$& $F\oplus H$ & $D\oplus E$ & $D\oplus E$ & $H\oplus C$ & $A\oplus B\oplus G$ & $F\oplus C$ \\ \hline
$\boldsymbol{H}$ &$H$ & $H$& $F\oplus G$ & $D\oplus E$ & $D\oplus E$ & $G\oplus C$ & $F\oplus C$ & $A\oplus B\oplus H$\\\hline
\end{tabular}
\end{table*}
}

\subsubsection{Algebraic theory}
\label{sec:ds3-algebraic-theory}

{\tiny{\tabulinesep=1.2mm
\begin{table*}\caption{Multi-fusion category $\mfC$ for the input category $\mC = \text{Vec}_{S_3}$. The diagonal entries are functor fusion categories $\mC_{ii} = \Fun_\mC(\M_i, \M_i)$. The off-diagonal entries are abelian semisimple categories $\mC_{ij} = \Fun_\mC(\M_i, \M_j)$; for these categories, we provide the quantum dimensions of the simple objects in curly braces.}
\label{tab:ds3-multi-fusion}
\centering
\begin{tabular}{|c|c|c|c|c|}
\hline
   $\boldsymbol{\mfC}$ & $\boldsymbol{\A_1}$ & $\boldsymbol{\A_2}$ & $\boldsymbol{\A_3}$ & $\boldsymbol{\A_4}$ \\ \hline
   $\boldsymbol{\A_1 = A+B+2C}$ / $\boldsymbol{K_1 = \{1\}}$ & $\text{Vec}_{S_3}$ & $\left\{\sqrt{3},\sqrt{3}\right\}$ & $\left\{\sqrt{2},\sqrt{2},\sqrt{2}\right\}$ & $\left\{\sqrt{6}\right\}$ \\ \hline
   $\boldsymbol{\A_2 = A+B+2F}$ / $\boldsymbol{K_2 = \Z_3}$ & $\left\{\sqrt{3},\sqrt{3}\right\}$ & $\text{Vec}_{S_3}$ & $\left\{\sqrt{6}\right\}$ & $\left\{\sqrt{2},\sqrt{2},\sqrt{2}\right\}$ \\ \hline
   $\boldsymbol{\A_3 = A+C+D}$ / $\boldsymbol{K_3 = \Z_2}$ & $\left\{\sqrt{2},\sqrt{2},\sqrt{2}\right\}$ & $\left\{\sqrt{6}\right\}$ & $\Rep(S_3)$ & $\left\{\sqrt{3},\sqrt{3}\right\}$ \\ \hline
   $\boldsymbol{\A_4 = A+F+D}$ / $\boldsymbol{K_4 = S_3}$ & $\left\{\sqrt{6}\right\}$ & $\left\{\sqrt{2},\sqrt{2},\sqrt{2}\right\}$ & $\left\{\sqrt{3},\sqrt{3}\right\}$ & $\Rep(S_3)$ \\ \hline
\end{tabular}
\end{table*}}}

As discussed in Refs. \onlinecite{Cong16a,Cong16b}, $\mfD(S_3)$ has four gapped boundary types, corresponding to the four subgroups of $S_3$ up to conjugation or the four Lagrangian algebras of $\mfD(S_3)$. These four Lagrangian algebras, along with all the corresponding fusion categories $\mC_{ii} = \Fun_\mC(\M_i, \M_i)$ and the data of the abelian semisimple categories $\mC_{ij}$ are shown in Table \ref{tab:ds3-multi-fusion}.

\subsubsection{The bulk-edge correspondence and braiding}
\label{sec:ds3-bulk-edge}

Let us now examine the bulk-edge correspondence between boundary defects and the $\Z_2$ symmetry defects of $\mfD(S_3)$. $\mfD(S_3)$ has one $\Z_2$ electric-magnetic symmetry given by interchanging the two topological charges $C$ and $F$. By Ref. \onlinecite{Barkeshli14}, the flux sector $\B_1$ contains four symmetry defects
of quantum dimension $\sqrt{3}$, and two of quantum dimension $2\sqrt{3}$. Our theory of crossed condensation then tells us that upon condensing to the gapped boundary $\A_1$ (resp. $\A_2$), the first four would condense to simple boundary defects of quantum dimension $\sqrt{3}$ in $\mC_{12}$ ($\mC_{21}$), while the last two would decompose as the direct sum of two such simple boundary defects. Similarly, the $C\leftrightarrow F$ symmetry between $\A_3$ and $\A_4$ allows us to say the same about condensing these symmetry defects onto the gapped boundaries $\A_3,\A_4$.

Through these bulk-edge correspondences and the results of Section \ref{sec:braiding}, we may also compute the projective braiding of boundary defects in $\mC_{12}$ with those in $\mC_{21}$, and similarly between boundary defects of $\mC_{34}$ and $\mC_{43}$. These braidings will be given by the $\Z_2$ crossed braiding of bulk symmetry defects. As discussed in Ref. \onlinecite{Barkeshli14}, fully gauging the $C \leftrightarrow F$ symmetry of $\mfD(S_3)$ yields the bilayer theory $\text{SU}(2)_4 \times \overbar{\text{SU}(2)_4}$, so we expect the projective braiding statistics of these boundary defects to be related to the braiding in the $\text{SU}(2)_4$ theory. Since the $\text{SU}(2)_4$ anyon theory has been made universal for topological quantum computation \cite{Cui15-m}, we believe these boundary defects could also be of importance for such purposes.

\subsubsection{Sequential condensation}

Through the above bulk-edge correspondence, we can now easily understand the boundary defects in $\mC_{12}$/$\mC_{21}$ and $\mC_{34}$/$\mC_{43}$ by considering the bulk $\Z_2$ symmetry defects associated with the $C \leftrightarrow F$ symmetry of $\mfD(S_3)$. We now present a mechanism---sequential condensation---which potentially allows us to understand some of the other cases.

When a condensible algebra $M$ has a condensible sub-algebra $N$, the condensation of $M$ can be done sequentially by condensing $N$ first.
In $\mfD(S_3)$, besides the Lagrangian algebras, there are also smaller condensible algebras such as $A+B, A+C$ and $A+F$. The quotient procedure of Section \ref{sec:condensation-tensor-functor} generalizes easily when we replace the Lagrangian algebra by a condensible sub-algebra. We first briefly explain how condensations of these smaller algebras work, and then use the results to obtain an intuitive understanding of boundary defects on defect sites.

Let us first consider the condensation of the  $A+B$ algebra. Since $B$ is the gauge charge of the $\mathbb{Z}_2$ subgroup of $S_3=\mathbb{Z}_3\rtimes \mathbb{Z}_2$, $\mfD(S_3)$ can be viewed as the gauging of a $\mathbb{Z}_2$ symmetry in the quantum double $\mfD(\mathbb{Z}_3)$, where the symmetry action can be canonically inferred from the semi-direct product structure. Therefore, this condensation results in $\mfD(\mathbb{Z}_3)$. In particular, $C$ and $F$ split into $e, \bar{e}=e^2$ and $m, \bar{m}=m^2$, respectively. The $\mathbb{Z}_2$ autoequivalence exchanging $C$ and $F$ becomes the one of $\mfD(\mathbb{Z}_3)$ exchanging $e$ and $m$.

For the $A+C$ algebra, one can show that the simple objects of the condensed theory are given by $A+C\; (\text{the vacuum}), B+C, D$ and $E$, all of which are abelian. We recognize that the resulting theory is simply $\mfD(\mathbb{Z}_2)$. By the $C\leftrightarrow F$ symmetry, we can obtain the condensation of $A+F$ as well.

Returning to Lagrangian algebras, we now consider $\A_1 = A+B+2C$. We can condense it in two steps: first condense $A+C$. In the condensed phase, we further condense the abelian boson corresponding to $B+C$, which then confines everything. In terms of the structure of the gapped boundary, we first form a domain wall between $\mfD(S_3)$ and $\mfD(\mathbb{Z}_2)$, followed by a gapped boundary between $\mfD(\mathbb{Z}_2)$ and the vacuum.

The condensation process of $A+B+2C$ or $A+B+2F$ can also be carried out as follows: first condense $A+B$, and then condense $e$ or $m$ in the resulting $\mfD(\mathbb{Z}_3)$ theory, corresponding to the algebra $A+B+2C$ and $A+B+2F$ in the original theory.

Condensation of $A+C+D$ can be done in a similar way. First we condense $A+C$ (notice that $A+D$ is not a condensible algebra), resulting in  $\mfD(\mathbb{Z}_2)$, then condensing the boson $D$ confines everything.

\begin{figure}[ht]
	\centering
	\includegraphics[width=0.62\columnwidth]{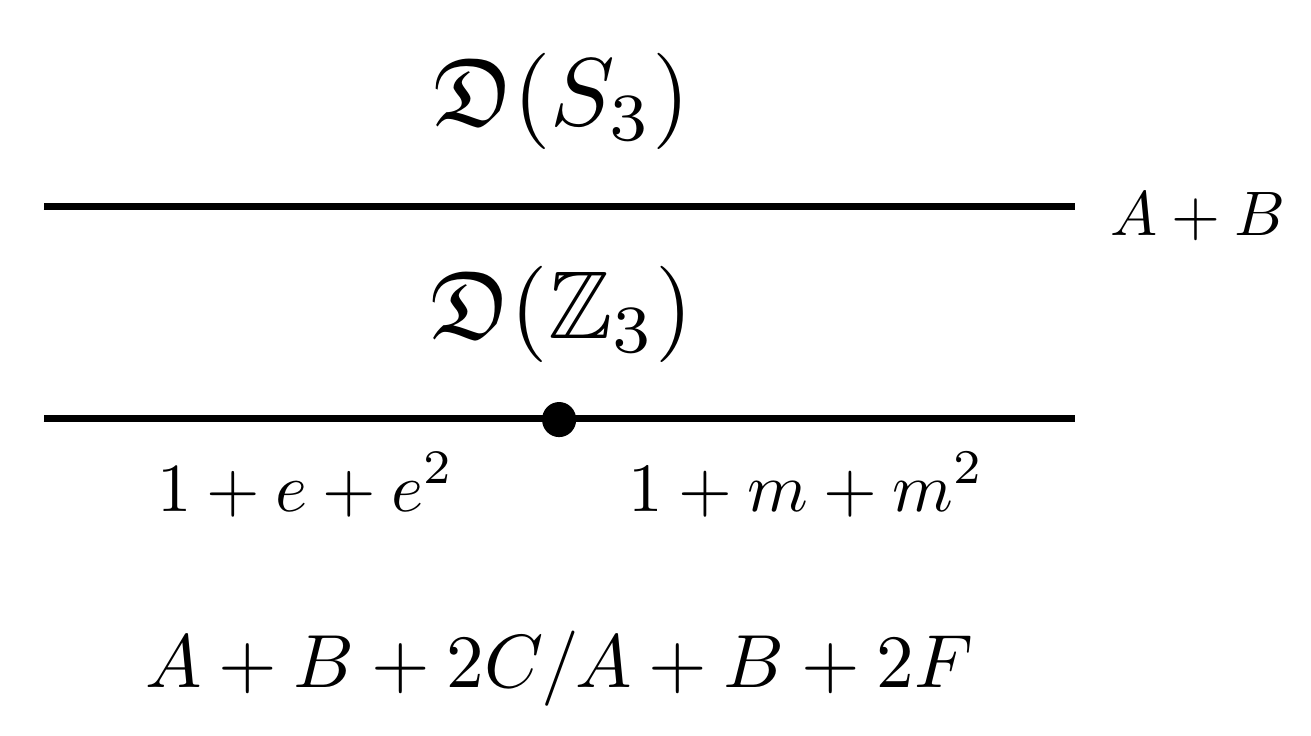}
	\caption{Illustration of the two-step construction for the $A+B+2C/A+B+2F$ boundary defects.}
	\label{fig:2step}
\end{figure}

This multi-step picture is particularly useful when we consider boundary defects between gapped boundaries. Take $A+B+2C$ and $A+B+2F$ as the first example. First condensing $A+B$, we obtain $\mfD(\mathbb{Z}_3)$. Next we condense $e$ or $m$ to form a boundary to vacuum. It is clear that the defect site between $A+B+2C$ and $A+B+2F$ is essentially the one between an $e$-type boundary and an $m$-type boundary in the $\mfD(\mathbb{Z}_3)$ intermediate layer. So we expect that these defect sites harbor a $\mathbb{Z}_3$ parafermion zero mode, which agrees with the quantum dimension and crossed condensation formulas. See Fig. \ref{fig:2step} for an illustration.

This method also allows us to go beyond the crossed condensation procedure, e.g. to explain the boundary defects between $A+B+2C$ and $A+C+D$. Here, we first condense $A+C$ to get $\mfD(\mathbb{Z}_2)$, and then the defect site is between an $e$-type edge (i.e $B+C$) and an $m$-type edge (i.e. $D$). So we expect $\mathbb{Z}_2$ Majorana zero mode to localize at the defect sites. Similar results are true for $A+B+2F$ and $A+D+F$ defect sites because of $C\leftrightarrow F$ symmetry in $\mfD(S_3)$. We see that these results agree with the quantum dimension formulas, although they are not directly explained by crossed condensation.

\subsubsection{Exact degeneracy}
\label{sec:ds3-exact-degeneracy}

Following the methods of Section \ref{sec:parafermion-exact-degeneracy}, we may also compute the exact degeneracy of boundary defects in the $\mfD(S_3)$ theory. Table \ref{ds3-exact-degeneracy} tallies the exact degeneracy obtained using two different methods: by exact diagonalization of the Hamiltonian with multiple boundary defects on a disk, and by counting fusion channels. In the following, we demonstrate the counting of fusion channels for several example cases.

\begin{table}\caption{Ground state degeneracy of $2n$ boundary defects on the edge of a disk, for gapped boundaries of $\mfD(S_3)$.}\label{ds3-exact-degeneracy}
\centering
\begin{tabular}{|c|c|c|c|c|}
\hline
   $\boldsymbol{\mfC}$ & $\boldsymbol{\A_1}$ & $\boldsymbol{\A_2}$ & $\boldsymbol{\A_3}$ & $\boldsymbol{\A_4}$ \\ \hline
   $\boldsymbol{\A_1}$ & $1$ & $3^{n-1}$ & $2^{n-1}$ & $6^{n-1}$ \\ \hline
   $\boldsymbol{\A_2}$  & $3^{n-1}$ & $1$ & $6^{n-1}$ & $2^{n-1}$ \\ \hline
   $\boldsymbol{\A_3}$ & $2^{n-1}$ & $6^{n-1}$ & $1$ & $\frac{1}{2}(3^{n-1}+1)$ \\ \hline
   $\boldsymbol{\A_4}$ & $6^{n-1}$ & $2^{n-1}$ & $\frac{1}{2}(3^{n-1}+1)$ & $1$ \\ \hline
\end{tabular}
\end{table}

\vspace{2mm}
\begin{itemize}[wide, label={}, listparindent=1.5em, parsep=0.25mm, itemsep=3mm, labelindent=0pt]
\item
{\it \underline{Case I}: Boundary defects between $\A_3 = A+C+D$ and $\A_4 = A+F+D$.}

In this case, the excitations on the $\A_3$ gapped boundary are given by 

\begin{equation}
\begin{gathered}
	1=(A+C+D), \quad X=(B+C+E),\\
	 Y=(D+E+F+G+H).
\end{gathered}
\end{equation}

\noindent
Notice that $d_X=1, d_Y=2$. Obviously, $X$ and $Y$ are confined, so they represent (solitonic) excitations on the boundary. The excitations $1, X$ and $Y$ form a fusion category with fusion rules of $\text{Rep}(S_3)$:

\begin{equation}
\begin{gathered}
	X\otimes X=1,\quad X\otimes Y=Y\otimes X=Y,\\
	Y\otimes Y=1+X+Y.
\end{gathered}
\end{equation}

To compute the fusion rule for $\tau_{34}\otimes \tau_{43}$, we apply our Conjecture \ref{zero-mode-conj}. We claim that the following fusion rule gives the correct degeneracy:
\begin{equation}
	\tau_{34}\otimes\tau_{43}=1+Y.
	\label{}
\end{equation}

Here, because $B$ is not a zero mode of $\tau_{34}$, $X$ does not appear in the right hand side. Another argument for this is the following: $X$ is morally the $B$ anyon, which has $-1$ braiding with $D$. So naively, one can imagine a Wilson line of $D$ terminating on the left and right of the $\mathcal{A}_4$ segment to measure whether the two boundary defects fuse into an $X$. However, we notice that $D$ is also condensed in $\mathcal{A}_4$. So one can say that in a sense there is no ``defect site'' for $D$. 

Let us now compute the degeneracy with the boundary defect fusion rule. Consider a configuration with $\mathcal{A}_3 - \mathcal{A}_4 - \mathcal{A}_3 - \mathcal{A}_4 -\cdots -\mathcal{A}_3$, with $2n$ boundary defects. Fusing all neighboring $\tau_{34}$ and $\tau_{43}$ results in $n$ excitations, each of which is either $1$ or $Y$. Now the question is to count the number of fusion trees with total charge $1$. We denote the number of fusion trees with a total charge $a$ by $D_n^a$, with the $n$ external lines being $1$ or $Y$. We find the following recursion relation:
\begin{equation}
	\begin{pmatrix}
		D_n^1\\
		D_n^X\\
		D_n^Y
	\end{pmatrix}=
	\begin{pmatrix}
		1 & 0 & 1\\
		0 & 1 & 1\\
		1 & 1 & 2\\
	\end{pmatrix}
\begin{pmatrix}
	D_{n-1}^1\\
	D_{n-1}^X\\
	D_{n-1}^Y
	\end{pmatrix}.
	\label{}
\end{equation}
\noindent
Since $D_2^1=2, D_2^X=1, D_2^Y=3$, we have 
\begin{equation}
	\begin{pmatrix}
		D_n^1\\
		D_n^X\\
		D_n^Y
	\end{pmatrix}=
	\begin{pmatrix}
		1 & 0 & 1\\
		0 & 1 & 1\\
		1 & 1 & 2\\
	\end{pmatrix}^{n-2}
	\begin{pmatrix}
		2\\
		1\\
		3
	\end{pmatrix}
\end{equation}
\noindent
and arrive at $D_n^1=\frac{1}{2}(3^{n-1}+1)$. Furthermore, $D_n^X=\frac{1}{2}(3^{n-1}-1), D_n^Y=3^{n-1}$.

\item
{\it \underline{Case II}: Boundary defects between $\A_3 = A+C+D$ and $\A_1 = A+B+2C$.}

Similarly, by examining the lifting and zero modes, we argue that 
\begin{equation}
	\tau_{31}\otimes\tau_{13}=1+X.
	\label{}
\end{equation}
This time, $Y$ does not appear because its lifting $F$ and $G$ are not zero modes of $\tau_{31}$. This fusion rule of course leads to the degeneracy $2^{n-1}$.

\item
{\it \underline{Case III}: Boundary defects between $\A_3 = A+C+D$ and $\A_2 = A+B+2F$.}

In this case, the lifting and zero modes give us the fusion rule
\begin{equation}
	\tau_{32}\otimes\tau_{23}=1+X+2Y.
	\label{}
\end{equation}
\noindent
The reason for the multiplicity $2$ in front of $Y$ is the following: recall that $Y=(D+E+F+G+H)$. For each bulk anyon in the lifting of $Y$, there are two topologically distinct ways to absorb it on the defect site. For example, consider bringing a $G$ (or $H$) anyon close to the defect site. We can split $G$ into $C\otimes F$ and condense $C$ and $F$ on the two sides of the defect site. Because the multiplicity $2$ of $F$ in the algebra $\mathcal{A}_2$, there are two ways that this process can happen. This is true for all anyons in the lifting of $Y$.

To compute the degeneracy, we first find the recursion relation:
\begin{equation}
	\begin{pmatrix}
		D_n^1\\
		D_n^X\\
		D_n^Y
	\end{pmatrix}=
	\begin{pmatrix}
		1 & 1 & 2\\
		1 & 1 & 2\\
		2 & 2 & 4\\
	\end{pmatrix}
\begin{pmatrix}
	D_{n-1}^1\\
	D_{n-1}^X\\
	D_{n-1}^Y
	\end{pmatrix}.
	\label{}
\end{equation}

\noindent
We have $D_2^1=6, D_2^X=6, D_2^Y=12$, so solving the recursion relation gives $D_n^1=6^{n-1}$, $D_n^X=6^{n-1}$, and  $D_n^Y=2\cdot 6^{n-1}$.
\end{itemize}

\vspace{2mm}
\section{Conclusions and Open Problems}
\label{sec:conclusions}

Motivated by the boundary defect realization of Majorana and parafermion zero modes, we developed a general exactly solvable Hamiltonian realization for boundary defects in Dijkgraaf-Witten theories based on any finite group $G$. By studying these boundary defects algebraically through a multi-fusion category, we developed a bulk-edge correspondence between symmetry defects in the bulk and certain defects between gapped boundaries, which also allowed us to obtain a projective braiding between the defects on the boundary. Finally, we considered several examples of importance to condensed matter physics and topological quantum computation.

We conclude by discussing a few potential generalizations of our work. First, while gapped boundaries between a topological phase $\B$ and the vacuum are mathematically equivalent to gapped domain walls between two different topological phases via a folding trick, it would still be interesting to study defects between different types of gapped domain walls as a generalization. Another direction is to generalize our Hamiltonian $H_{\text{dft}}$ to the twisted Dijkgraaf-Witten model, and later to the Levin-Wen model using quantum groupoids.

In general, given an abelian topological phase of matter, two prominent ways to generate non-abelian objects are (1) to introduce gapped boundaries, and (2) to introduce symmetry defects. Our analysis of boundary defects and the bulk-edge correspondence gives a way to introduce both together, and elucidates a precise relationship between them. We believe this would be especially important for applications to topological quantum computation. An interesting question here would be to find universal gate sets and their physical implementations for qudits based on degeneracy of many boundary defects, or more generally, to find such gate sets based on the combined degeneracy of gapped boundaries, boundary defects, and symmetry defects. One possible way may be to obtain surface code implementations of boundary defects in general Dijkgraaf-Witten theories (beyond $G = \Z_p$) using our Hamiltonian $H_{\text{dft}}$.

While the degeneracy of many boundary defects can be counted through the fusion channels, it would be interesting to understand the degeneracy using ribbon operators.  Such an understanding would shed light on the projective braid statistics of boundary defects, and point to potential physical realizations of quantum gates.



\begin{thebibliography}{1}%
\makeatletter
\providecommand \@ifxundefined [1]{%
 \@ifx{#1\undefined}
}%
\providecommand \@ifnum [1]{%
 \ifnum #1\expandafter \@firstoftwo
 \else \expandafter \@secondoftwo
 \fi
}%
\providecommand \@ifx [1]{%
 \ifx #1\expandafter \@firstoftwo
 \else \expandafter \@secondoftwo
 \fi
}%
\providecommand \natexlab [1]{#1}%
\providecommand \enquote  [1]{``#1''}%
\providecommand \bibnamefont  [1]{#1}%
\providecommand \bibfnamefont [1]{#1}%
\providecommand \citenamefont [1]{#1}%
\providecommand \href@noop [0]{\@secondoftwo}%
\providecommand \href [0]{\begingroup \@sanitize@url \@href}%
\providecommand \@href[1]{\@@startlink{#1}\@@href}%
\providecommand \@@href[1]{\endgroup#1\@@endlink}%
\providecommand \@sanitize@url [0]{\catcode `\\12\catcode `\$12\catcode
  `\&12\catcode `\#12\catcode `\^12\catcode `\_12\catcode `\%12\relax}%
\providecommand \@@startlink[1]{}%
\providecommand \@@endlink[0]{}%
\providecommand \url  [0]{\begingroup\@sanitize@url \@url }%
\providecommand \@url [1]{\endgroup\@href {#1}{\urlprefix }}%
\providecommand \urlprefix  [0]{URL }%
\providecommand \Eprint [0]{\href }%
\providecommand \doibase [0]{http://dx.doi.org/}%
\providecommand \selectlanguage [0]{\@gobble}%
\providecommand \bibinfo  [0]{\@secondoftwo}%
\providecommand \bibfield  [0]{\@secondoftwo}%
\providecommand \translation [1]{[#1]}%
\providecommand \BibitemOpen [0]{}%
\providecommand \bibitemStop [0]{}%
\providecommand \bibitemNoStop [0]{.\EOS\space}%
\providecommand \EOS [0]{\spacefactor3000\relax}%
\providecommand \BibitemShut  [1]{\csname bibitem#1\endcsname}%
\let\auto@bib@innerbib\@empty
\bibitem [{Note1()}]{Note1}%
  \BibitemOpen
  \bibinfo {note} {We use the notation $\protect \text {FPdim}$ to denote the
  quantum dimension of the anyon corresponding to the label $(T,R)$, to
  contrast with the notation $\protect \Dim $ for the usual dimension of a
  representation.}\BibitemShut {Stop}%
\end{thebibliography}%


\clearpage

\vspace{4mm}

\begin{thebibliography}{99}
\bibliographystyle{alpha}
\bibitem{BakalovKirillov}
B. Bakalov, A. A. Kirillov. {\it Lectures on tensor categories and modular functors.} Vol. 21. American Mathematical Soc. (2001).
\bibitem{Barkeshli11}
M. Barkeshli and X.-L. Qi. {\it Topological Nematic States and Non-Abelian Lattice Dislocations}. Phys. Rev. X {\bf 2}(3), 031013 (2012).
\bibitem{Barkeshli14}
M. Barkeshli, P. Bonderson, M. Cheng, Z. Wang. {\it Symmetry, defects, and gauging of topological phases}. arXiv preprint arXiv:1410.4540 (2014).
\bibitem{Barkeshli15}
M. Barkeshli and J. D. Sau, {\it Physical Architecture for a Universal Topological Quantum Computer based on a Network of Majorana Nanowires}. arXiv preprint arxiv:1509.07135 (2015).
\bibitem{Barkeshli16}
M. Barkeshli, M. Freedman. {\it Modular transformations through sequences of topological charge projections}. arXiv preprint arXiv:1602.01093 (2016).
\bibitem{Bark13a}M. Barkeshli, C.-M. Jian, X.-L. Qi. {\it Twist defects and projective non-Abelian braiding statistics}. Phys. Rev. B {\bf 87}(4), 045130 (2013).
\bibitem{Bark13b}M. Barkeshli, C.-M. Jian, X.-L. Qi. {\it Theory of defects in Abelian topological states}. Phys. Rev.B {\bf 88}(23), 235103 (2013).
\bibitem{Bark13c}M. Barkeshli, C.-M. Jian,  X.-L. Qi. {\it Classification of topological defects in Abelian topological states}. Phys. Rev. B {\bf 88}(24), 241103 (2013).
\bibitem{BK+14}Barends, Rami, et al. {\it Superconducting quantum circuits at the surface code threshold for fault tolerance.} Nature {\bf 508}(7497), 500-503 (2014).
\bibitem{Beigi11}
S. Beigi, P. W. Shor, D. Whalen. {\it The quantum double model with boundary: condensations and symmetries}. Communications in Mathematical Physics {\bf 306}(3), 663-694 (2011).
\bibitem{Bombin08}
H. Bombin, M. A. Martin-Delgado. {\it Family of non-Abelian Kitaev models on a lattice: Topological condensation and confinement}. Phys. Rev. B {\bf 78}, 115421 (2008).
\bibitem{Bombin11}
H. Bombin, M. A. Martin-Delgado. {\it Nested Topological Order}. New J. Phys. {\bf 13}, 125001 (2011).
\bibitem{Bonderson13}
P. Bonderson. {\it Measurement-Only Topological Quantum Computation via Tunable Interactions}. Phys. Rev. B {\bf 87}, 035113 (2013).
\bibitem{Bravyi98} 
S. B. Bravyi, A. Y. Kitaev. {\it Quantum codes on a lattice with boundary}. arXiv:quant-ph/9811052 (1998).
\bibitem{Brown16}
B. J. Brown, K. Laubscher, M. K. Kesselring, and J. R. Wootton. {\it Poking holes and cutting corners to achieve Clifford gates with the surface code}. arXiv preprint arXiv:1609.04673v3 (2016).
\bibitem{Chang15}
L. Chang, et al. {\it On enriching the Levin-Wen model with symmetry.} J. Phys. A: Mathematical and Theoretical, {\bf 48}(12), 12FT01 (2015).
\bibitem{cheng2012}
M. Cheng. Phys. Rev. B, {\it Superconducting Proximity Effect on the Edge of Fractional Topological Insulators}. {\bf 86}, 195126 (2012).
\bibitem{clarke2013}  
D.~J. Clarke, J. Alicea and K. Shtengel.  {\it Exotic non-Abelian anyons from conventional fractional quantum Hall states}. {Nature Comm.} {\bf 4}, 1348 (2013).
\bibitem{Cong16a}
I. Cong, M. Cheng, Z. Wang. {\it Topological quantum computation with gapped boundaries.} arXiv preprint arXiv:1609.02037 (2016).
\bibitem{Cong16b}
I. Cong, M. Cheng, Z. Wang. {\it Hamiltonian and Algebraic Theories of Gapped Boundaries in Topological Phases of Matter}. Submitted (2016).
\bibitem{Cui15-m}
S. X. Cui, Z. Wang. {\it Universal quantum computation with metaplectic anyons}. J. Math. Phys. {\bf 56}, 032202 (2015).
\bibitem{Cui15}
S. X. Cui, S.-M. Hong, Z. Wang. {\it Universal quantum computation with weakly integral anyons}. Quantum Inf. Process. 14:2687–2727 (2015).
\bibitem{DFN}S. Das Sarma, M. Freedman, C. Nayak. {\it Majorana zero modes and topological quantum computation.} arXiv preprint arXiv:1501.02813 (2015).
\bibitem{Davydov12}
A. Davydov, M. M{\"u}ger, D. Nikshych, V. Ostrik. {\it The Witt group of non-degenerate braided fusion categories.} Journal f{\"u}r die reine und angewandte Mathematik, {\bf 677} 177 (2012).
\bibitem{Etingof10}
P. Etingof, D. Nikshych, V. Ostrik. {\it Fusion categories and homotopy theory}, arXiv:0909.3140 (2010).
\bibitem{Etingof15}
P. Etingof, S. Gelaki, D. Nikshych, V. Ostrik. {\it Tensor categories}. Vol. 2015. American Mathematical Society (2015).
\bibitem{Fowler12}
A.G. Fowler, M. Mariantoni, J. M. Martinis, A. N. Cleland. {\it Surface codes: Towards practical large-scale quantum computation}. Phys. Rev. A {\bf 86}(3), 032324 (2012).
\bibitem{FuKa}L. Fu, C. L. Kane. {\it Superconducting proximity effect and Majorana fermions at the surface of a topological insulator.} Phys. Rev. Lett. {\bf 100}(9), 096407 (2008).
\bibitem{HNW}M.-B Hastings, C. Nayak, Z. Wang. {\it Metaplectic anyons, Majorana zero modes, and their computational power.} Phys. Rev. B {\bf 87}(16) (2013): 165421.
\bibitem{Hung15a}
L.-Y. Hung, Y. Wan. {\it Generalized ADE classification of topological boundaries and anyon condensation}, Journal of High Energy Physics 2015 (07), 120 (2015).
\bibitem{Hung15b}
L.-Y. Hung, Y. Wan. {\it Ground-State Degeneracy of Topological Phases on Open Surfaces}, Phys. Rev. Lett. {\bf 114}, 076401 (2015).
\bibitem{Kapustin14}
A. Kapustin, {\it Ground-state degeneracy for Abelian anyons in the presence of gapped boundaries}, Phys. Rev. B {\bf 89}, 125307 (2014).
\bibitem{Kitaev97}
A. Y. Kitaev. {\it Fault-tolerant quantum computation by anyons}. Ann. Phys. {\bf 303}(2) (2003).
\bibitem{KitaevKong}
A. Kitaev, L. Kong, {\it Models for Gapped Boundaries and Domain Walls}. Commun. Math. Phys. {\bf 313}, 351-373 (2012). doi: 10.1007/s00220-012-1500-5.
\bibitem{Kong}L. Kong. {\it Some universal properties of Levin-Wen models}. XVIITH International Congress of Mathematical Physics, World Scientific (2014).
\bibitem{Kong15}
L. Kong, X.-G. Wen, Z. Hao, {\it Boundary-bulk relation for topological orders as the functor mapping higher categories to their centers}. arXiv:1502.01690 (2015)
\bibitem{LWW}T. Lan, J. C. Wang, X.-G. Wen. {\it Gapped Domain Walls, Gapped Boundaries, and Topological Degeneracy}. Phys. Rev. Lett. {\bf 114}(7), 076402 (2015).
\bibitem{Levin04}
M. A. Levin, X.-G. Wen. {\it String-net condensation: A physical mechanism for topological
phases}. Phys. Rev. B {\bf 71}, 045110 (2005)
\bibitem{Levin13}
M. Levin, {\it Protected edge modes without symmetry}. Phys. Rev. X {\bf 3}, 021009 (2013).
\bibitem{lindner2012}
N.~H. Lindner, E. Berg, G. Refael and A. Stern. {\it Fractionalizing Majorana fermions: non-abelian statistics on the edges of abelian quantum Hall states}. Phys. Rev. X, {\bf 2}, 041002 (2012).
\bibitem{MacLane}
S. Mac Lane. {\it Categories for the working mathematician.} Vol. 5. Springer Science and Business Media, 2013.
\bibitem{Muger03}
M. M{\"u}ger. {\it Galois extensions of braided tensor categories and braided crossed G-categories}. Journal of Algebra {\bf 277} 256-281 (2004).
\bibitem{Ostrik02}
V. Ostrik. {\it Module categories over the Drinfeld double of a finite group.} Int. Math (2002).
\bibitem{WW}J.C. Wang, X.-G. Wen. {\it Boundary degeneracy of topological order}.Phys. Rev. B. {\bf 91}(12), 125124 (2015).
\bibitem{Yamagami}S. Yamagami, {\it Group symmetry in tensor categories and duality for orbifolds},
Journal of Pure and Applied Algebra, {\bf 167}(1), 83-128 (2002).
\bibitem{Note1}
Our Hamiltonian generalizes easily to arbitrary lattices; we choose the square one for simplicity of presentation and calculation.
\bibitem{Note2}
The particular choice of these two cilia is due to the choice of orienting the boundary counterclockwise, as discussed in Section \ref{sec:hamiltonian-background}; if we oriented the boundary clockwise, the defects would be located at the mirror images across $\mfL$ of the highlighted cilia.
\bibitem{Note3}
This is a conflict of notation with the $G$ of Sec. \ref{sec:hamiltonian-realization} which is used for the input group for Kitaev's quantum double models. We use $G$ for the symmetry group here as well to adhere with existing literature (e.g. Ref. \onlinecite{Barkeshli14}); the distinction is mostly clear.
\bibitem{Note4}
Since $\B$ is braided, any left or right module category becomes a bi-module category through the braiding.
\bibitem{Note5}
Figure credit to Ref. \onlinecite{Barkeshli14}.
\end{thebibliography}
\end{document}